\documentclass{aa}
\usepackage[utf8]{inputenc}
\usepackage[colorlinks]{hyperref}
\usepackage{amsmath}
\usepackage{amssymb}
\usepackage{microtype}
\usepackage[T1]{fontenc}
\usepackage{lmodern}
\usepackage{natbib}
\usepackage{graphicx}
\usepackage{microtype}
\usepackage{color}
\usepackage[dvipsnames]{xcolor}
\usepackage{multirow}
\usepackage{xspace}
\usepackage[normalem]{ulem}
\graphicspath{ {figures/} }

\usepackage{pdflscape} 

\bibpunct{(}{)}{;}{a}{}{,} 

\def\aap{A\&A}

\def\apjs{ApJS}

\def\apjl{ApJLett}

\def\srg{\textit{SRG}}
\def\art{ART-XC}
\def\erosita{eROSITA}
\def\uhuru{\textit{Uhuru}}
\def\rosat{\textit{ROSAT}}
\def\rxte{\textit{RXTE}}
\def\maxi{MAXI}
\def\integral{\textit{INTEGRAL}}
\def\swift{\textit{Swift}}

\def\xmm{\textit{XMM-Newton}}
\def\fermi{\textit{Fermi}}

\def\arcmin{\hbox{$^\prime$}}
\def\arcsec{\hbox{$^{\prime\prime}$}}
\def\flux{erg\,s$^{-1}$\,cm$^{-2}$}

\def\Nf{\ensuremath{N_{\rm f}}}
\def\Pf{\ensuremath{P_{\rm f}}}

\begin{document}

\title{\srg/\art\ all-sky X-ray survey: Catalog of sources detected during the first year\thanks{The catalog is only available in electronic form at the CDS via anonymous ftp to cdsarc.u-strasbg.fr (130.79.128.5) or via \url{http://cdsweb.u-strasbg.fr/cgi-bin/qcat?J/A+A/} and at \url{http://srg.cosmos.ru}.}}

\author{M.~Pavlinsky\inst{1} \and S. Sazonov\inst{1}\thanks{E-mail: sazonov@iki.rssi.ru} \and R. Burenin\inst{1} \and E. Filippova\inst{1} \and R.~Krivonos\inst{1} \and V. Arefiev\inst{1} \and M. Buntov\inst{1} \and C.-T. Chen\inst{2} \and S. Ehlert\inst{3} \and I. Lapshov\inst{1} \and V. Levin\inst{1} \and A. Lutovinov\inst{1} \and A. Lyapin\inst{1} \and I. Mereminskiy\inst{1} \and S. Molkov\inst{1} \and B.~D.~Ramsey\inst{3} \and A. Semena\inst{1} \and N. Semena\inst{1} \and A. Shtykovsky\inst{1} \and R. Sunyaev\inst{1} \and A. Tkachenko\inst{1} \and D. A. Swartz\inst{2} \and A. Vikhlinin\inst{1,4}
}

\institute{
    Space Research Institute, 84/32 Profsouznaya str., Moscow 117997, Russian Federation
    \and Universities Space Research Association, Huntsville, AL 35805, USA
    \and NASA/Marshall Space Flight Center, Huntsville, AL 35812 USA
    \and Harvard-Smithsonian Center for Astrophysics, 60 Garden Street, Cambridge, MA 02138, USA
}

\abstract{
We present a first catalog of sources detected by the Mikhail Pavlinsky \art\ telescope on board the \srg\ observatory in the 4--12\,keV energy band during its ongoing all-sky survey. The catalog comprises 867 sources detected on the combined map of the first two 6-month scans of the sky (December 2019 -- December 2020), \art\ sky surveys 1 and 2, or ARTSS12. The achieved sensitivity to point sources varies between $\sim 4\times 10^{-12}$\,\flux\ near the ecliptic plane and $\sim 8\times 10^{-13}$\,\flux\ (4--12\,keV) near the ecliptic poles, and the typical localization accuracy is $\sim 15$\arcsec. Of the 750 sources of known or suspected origin in the catalog, 56\% are extragalactic (mostly active galactic nuclei, AGN; and clusters of galaxies) and the rest are Galactic (mostly cataclysmic variables, CVs; and low- and high-mass X-ray binaries). For 114 sources, \art\ has detected X-rays for the first time. Although the majority of these ($\sim 80$) are expected to be spurious (given the adopted detection threshold), there can be a significant number of newly discovered astrophysical objects. We have started a program of optical follow-up observations of the new and previously unidentified X-ray sources, which has already led to the identification of several AGN and CVs. With the \srg\ all-sky survey planned to continue for a total of four\,years, we can expect the \art\ survey in the 4--12\,keV band to significantly surpass previous surveys that were carried out in similar (medium X-ray) energy bands in terms of the combination of angular resolution, sensitivity, and sky coverage. 
}

\authorrunning{}

\maketitle

\section{Introduction}

All-sky X-ray surveys have played a paramount role in the exploration of Galactic and extragalactic astrophysical objects ever since the first such survey was carried out by the \uhuru\ orbital observatory in 1970--1973 \citep{Forman78}. A revolutionary all-sky survey in soft X-rays was performed in 1990--1991 by the {\it R\"ontgensatellit} (\rosat) orbital observatory. The grazing incidence X-ray telescope on \rosat\ improved the sensitivity by two orders of magnitude with respect to previous surveys, which enabled the detection of $\sim 10^5$ sources in the 0.1--2.4\,keV energy band \citep{Voges99,Boller16}. 

However, soft X-rays provide a strongly biased view of the Universe because X-ray sources are frequently obscured from our view by gas and dust within the object and/or in the intervening matter. Hence, there is a need for large-area surveys that are performed at energies above a few keV. Since the beginning of the 21st century, a number of such surveys have been performed in the standard (2--10\,keV) and harder X-ray energy bands. Most of them are serendipitous, that is, they are based on a compilation of pointed or slew observations. The {\it Rossi X-ray Timing Explorer} (\rxte) Slew Survey (XSS) achieved a sensitivity $\sim 10^{-11}$\,\flux\ and angular resolution $\sim 1^\circ$ in the 3--20\,keV energy band in the extragalactic sky ($|b|>10^\circ$) \citep{Revnivtsev04,Sazonov04}. A somewhat better sensitivity ($\sim 5\times 10^{-12}$\,\flux) was recently reached in the 4--10\,keV band in the Monitor of All-sky X-ray Image/Gas Slit Camera (\maxi/GSC) all-sky survey \citep{Kawamuro18,Hori18}, which similarly to XSS is conducted by a collimator instrument. A similar average depth in the 2--12\,keV band, but with excellent angular resolution (achieved through X-ray mirror optics), characterizes the \xmm\ Slew Survey (XMMSL, \citealt{Saxton08}); however, XMMSL has covered the sky highly nonuniformly and not in its entirety (84\% by the end of 2014, the time of compilation of the XMMSL2.0 catalog\footnote{https://www.cosmos.esa.int/web/xmm-newton/xmmsl2-ug}). Finally, hard X-ray (above 15\,keV) all-sky surveys carried out by the coded-mask Imager on-Board the INTEGRAL Satellite (IBIS) on board the {\it INTErnational Gamma-Ray Astrophysics Laboratory} (\integral) and the Burst Alert Telescope (BAT) on board the Neil Gehrels \swift\ observatory have reached a depth $\sim 10^{-11}$\,\flux\ and angular resolution $\sim 10$\,arcmin (e.g., \citealt{Krivonos2007,Krivonos12,Bird16,Oh18}). All these surveys together have discovered several hundred new X-ray sources that had not been detected in soft X-rays during the \rosat\ all-sky survey, in particular, a large number of heavily obscured active galactic nuclei (AGN) and high-mass X-ray binaries (HMXBs). These discoveries have greatly improved our understanding of the corresponding populations of objects in the local Universe and in the Galaxy (e.g., \citealt{2017ApJ...850...74K,Kretschmar2019,Malizia2020}). 

The {\it Spektrum-Roentgen-Gamma} (\srg) orbital observatory\footnote{\url{http://srg.cosmos.ru}} \citep{Sunyaev21}, launched on 13 July 2019 from the Baikonur Cosmodrome to a halo orbit near the L2 point of the Sun--Earth system, promises to once again revolutionize our understanding of various populations of X-ray sources. The observatory is equipped with two grazing incidence telescopes: the extended ROentgen Survey with an Imaging Telescope Array (\erosita) \citep{Predehl21}, and the Mikhail Pavlinsky Astronomical Roentgen Telescope -- X-ray Concentrator (\art) \citep{Pavlinsky2021}, which operate in the overlapping 0.2--8\,keV and 4--30\,keV energy bands, respectively. Since 12 December 2019, \srg\ has been performing an all-sky X-ray survey, which is planned to consist of eight scans of the whole sky, each lasting 6 months. 

At energies above 2\,keV, \erosita\ and \art\ will for the first time survey the whole sky with subarcminute angular resolution. The \art\ survey in the 4--12\,keV band is expected to significantly surpass previous surveys that were carried out in similar energy bands in terms of the combination of angular resolution, sensitivity, and uniformity. \art\ is a crucial component of the \srg\ mission because it provides better sensitivity than \erosita\ at energies above 6\,keV \citep{Sunyaev21}. \art\ is particularly important for systematic search and exploration of absorbed X-ray sources.

By December 15, 2020, \srg\ had completed two full scans of the sky. This paper presents a catalog of sources detected by \art\ during this early stage of the mission. 

\section{Data analysis}
\label{s:data}

After an initial calibration and performance verification (CalPV) phase, the \srg\ observatory started conducting its all-sky X-ray survey on 12 December 2019. During the survey, the optical axes of the \art\ and \erosita\ telescopes rotate with a period of 4\,hours around the spacecraft Z-axis that is pointed approximately toward the Sun (this main regime of observations is hereafter referred to as ``survey mode''). This enables full sky coverage every 6\,months (see \citealt{Sunyaev21} for further details). By 10 June 2020, the entire sky had been covered for the first time and by 15 December 2020 for the second time. The catalog of sources presented in this paper is based on the \art\ data accumulated during these two surveys, hereafter referred to as \art\ sky survey 1 and \art\ sky survey 2 (or ARTSS1 and ARTSS2 for short), respectively. 

We only used the \art\ data obtained in survey mode and disregarded the data obtained during deep observations of selected regions of the sky in scanning mode (when the \art\ and \erosita\ telescopes perform a raster scan of a field with a size up to $12.5^\circ \times 12.5^\circ$, see \citealt{Sunyaev21,Pavlinsky2021} for further details) and during pointed observations\footnote{Scanning and pointed observations have taken place not only during the CalPV phase, but also during short recesses in the all-sky survey associated with \srg\ orbit corrections.}. Time intervals when the calibration sources were inserted into the collimators or when high voltage was switched off on the detectors for depolarization (see \citealt{Pavlinsky2021} for details) were also excised. As the \art\ background has proved to be exceptionally stable \citep{Pavlinsky2021}, no cleaning of \art\ X-ray data for high background periods was needed. Only events detected in one or two upper detector strips and in one or two lower strips were selected, while events detected in a larger number of strips were not used in the analysis\footnote{The coordinate resolution of each of the seven \art\ telescope modules is provided by two mutually perpendicular sets of 48 strips on the two sides of a CdTe crystal. The strip width corresponds to an angular resolution of 45\arcsec\ \citep{Pavlinsky2021}.}.

We produced a summed map of ARTSS1 and ARTSS2 (the combined first-year survey is hereafter referred to as ARTSS12) in the 4--12\,keV energy band. The choice of this band is motivated by the energy dependence of the \art\ effective area (see fig.~19 in \citealt{Pavlinsky2021}), namely by a significant and abrupt drop in sensitivity above 12\,keV. We then applied to this map a set of algorithms for detection of point and extended X-ray sources that are described in \S\ref{s:detpoint} and \S\ref{s:detext}, respectively, which resulted in a catalog of sources detected during ARTSS12. 

\subsection{Construction of maps}
\label{s:maps}

In order to construct all-sky maps, the \art\ survey data were split into 4,700 overlapping sky tiles in equatorial coordinates. All tiles have an equal tangential size of $3.6 \times 3.6$\,degrees, and the overlap between tiles is 0.6\,degree in the right ascension and declination directions. This is needed to eliminate edge effects during sky background determination and source detection. The same scheme is used for \erosita\ data products \citep{Predehl21}. For each tile, a set of standard maps were prepared, including an exposure map, particle and photon background maps, and sky images. All maps consist of $1024\times 1024$ pixels of 12.66\arcsec\ . The maps were prepared for ARTSS1 and ARTSS2 separately and were then combined.

When combining the data of the seven \art\ telescope modules, we took the misalignment of their optical axes and the actual positions of the optical axes at the detectors into account, which had been calibrated using observations of bright X-ray sources during the survey and a special series of pointed observations of the Crab nebula, respectively. The exposure maps were corrected for the vignetting function, which had been measured using ground calibrations of the mirrors and detector units as well as extensive ray-tracing simulations. We refer to \cite{Pavlinsky2021}, where all the instrument calibration activities preceding the construction of maps are discussed in detail.

\subsection{Background estimation}
\label{s:bgr}

To estimate the particle background, we took advantage of the fact that the efficiency of the \art\ X-ray optics vanishes at energies higher than 30\,keV, and we used observations of blank-sky fields (i.e., regions without bright X-ray sources). Assuming that all events in these blank sky observations are associated with particle background, we determined the ratio of count rates in the energy bands of 4--12\,keV and 30--70\,keV for each detector pixel. These ratios were then combined with the measured 30--70\,keV count rate for each survey mode observation to determine the expected number of particle background events in the 4--12\,keV energy band in each detector pixel during each 1\,s time interval of the survey. These values were then projected onto the sky map. Because the particle background measured in the \art\ orbit is extremely stable, we expect this background estimation to be robust. The cosmic X-ray background provides a negligible contribution to the total background in the 4--12\,keV energy band.

The residual background, associated with the uncertainty of the particle background estimation and with possible large-scale X-ray structures in the sky (e.g., the Galactic Ridge X-ray emission), was estimated from X-ray images themselves. To this end, the particle background was subtracted from a given X-ray image and then all significant details at angular scales below 5\arcmin\ are eliminated from the X-ray image using the wavelet decomposition technique (\texttt{wvdecomp}, \citealt{1998ApJ...502..558V}). 
This angular scale was chosen so that it is much larger than the full width at half maximum (FWHM) of the \art\ point spread function (PSF), equal to 53\arcsec\ \citep{Pavlinsky2021}, but is much smaller than the \art\ field of view (FoV), 36\arcmin\ in diameter \citep{Pavlinsky2021}.

\subsection{Detection of point sources}
\label{s:detpoint}

Convolution of an image with an instrument's PSF is equivalent to a maximum likelihood test for the case of high background when the statistics is Gaussian (e.g., \citealt{Pratt1978}). However, in X-ray images, the noise statistics is often Poisson, and the optimal filter, which maximizes the probability of detection of real sources and minimizes the probability of false detections of statistical fluctuations,  can in this case be written as \citep{Lynx_CSR_2018,Ofek18}
\begin{equation}\label{filter}
\Phi=\ln \left(\frac{f}{b} P + 1 \right),
\end{equation}
where $f$ is the source flux, $b$ is the background brightness, and $P$ is the PSF image normalized to unity. This approach can be readily extended to the case of any specific spectral flux distribution.

In the case of \art\ data, the situation is even more complicated because the PSF is not centrally symmetric and the PSF and vignetting strongly vary across the FoV. Therefore we used the following filter for source detection:
\begin{equation}
  \Phi(x) = \ln \left( \frac{f(e) v(x,e)}{b(x,e)} P(x_0 | x) + 1 \right),
  \label{eq:art_phi_x}
\end{equation}
where $x_0$ is the photon coordinates, $e$ is the photon energy, $f(e)$ is the expected spectral energy distribution of X-ray sources, $v(x,e)$ is the vignetting function, $b(x,e)$ is the spectral brightness of the background as a function of coordinates and energy, and $P(x_0 | x)$ is the PSF value at $x_0$ under the assumption that the source is located at $x$. The optimal filter described by equation~(\ref{eq:art_phi_x}) thus depends on the energy of every photon and on the estimated background in every point in the image. Hence, it is applied to \art\ data on an event-by-event basis.

We adopted the vignetting function from the results of extensive ray-tracing simulations of the \art\ X-ray mirrors \citep{Pavlinsky2021}. The $P(x_0 | x)$ function was precomputed for a fine grid of positions in the \art\ FoV using the \art\ PSF model obtained from the results of ground calibrations of the mirror systems \citep{2014SPIE.9144E..4UG,2017ExA....44..147K,2018ExA....45..315P,2019ExA....47....1P,2019ExA....48..233P,Pavlinsky2021}.
Interpolation was then used to evaluate $P(x_0 | x)$ for a photon at a given position in the \art\ FoV.

For X-ray sources, we assumed a power-law spectrum with a photon index $\Gamma=1.4$, which approximately corresponds to the slope of the cosmic X-ray background spectrum in the 4--20\,keV energy band considered here \citep{Gruber1999,Churazov2007,Krivonos2021} and to the slope of the cumulative X-ray spectrum of the local population of AGN \citep{Sazonov2008}. This is a reasonable assumption (which can be improved in future analyses of \art\ all-sky survey data) because nearly half of the ARTSS12 sources prove to be AGN, whereas the remaining ones are mostly X-ray binaries and cataclysmic variables (CVs), whose X-ray spectra are not dramatically different from those of AGN. For the background, we adopted a flat spectrum with a photon index $\Gamma=0$, which is close to the measured \art\ particle background spectrum in the 4--20\,keV energy band. 

To calculate the flux-to-background ratio in equation~(\ref{eq:art_phi_x}), we assumed that point sources have a diameter of 53\arcsec, that is, that the diameter is equal to the FWHM of the \art\ PSF. Given the \art\ PSF, this size is also the diameter of the circle within which half of the source photons with angular offsets smaller than 5\arcmin\ are contained (source photons with larger offsets are included in the background, see \S\ref{s:bgr} above).

Sources are detected in images that are obtained by convolution of raw \art\ data with the optimal filter given by equation~(\ref{eq:art_phi_x}) (hereafter referred to as convolution images) as peaks with values above some specified threshold. In the case of multiple detections within 53\arcsec\ from each other, only the highest peak is taken into account (with the corresponding coordinates and amplitude).


In order to specify thresholds for source detection in convolution images, we carried out Monte Carlo simulations of empty fields (Burenin et al., in preparation). Only the particle background was simulated because it strongly dominates the \art\ background. We thus computed the expected number of peaks of a given amplitude per square degree in the convolution image of an empty field, $\Nf$, for different background levels. Using these tabulated results of simulations, we can evaluate the probability for a source detection in a given \art\ convolution image (with the corresponding background level) to be spurious as $\Pf = \Nf/N_{\rm tr}$, where $N_{\rm tr}$ is the effective number of trials. The latter can be estimated from the adopted size of pointed sources (53\arcsec) as the number of independent cells for source detection per square degree. Finally, the statistical significance (signal-to-noise ratio), $S/N$, of the source detection can be determined from the likelihood $L = -2 \ln \Pf + \mathrm{const}$, where the constant is defined so that the value of the complimentary error function at $S/N = \sqrt{L}$ is equal to $\Pf$.

We adopted the $S/N=4.5$ level as a threshold for detection of pointed sources. The source flux
within the 53\arcsec\ aperture that corresponds to $S/N=4.5$ for a given
background level must then be substituted into equation~(\ref{eq:art_phi_x}). The resulting filter is optimal for the detection of weak sources near the adopted $S/N$ threshold, with the corresponding flux threshold depending on the estimated background at a given position in the sky. However, brighter sources will also be confidently detected with this filter (see section~A.3.2 in \citealt{Lynx_CSR_2018}). Eventually (in \S\ref{s:thresh}), we imposed a slightly (by $\sim 7$\%) higher $S/N$ cut to include sources into a final \art\ catalog to reduce the fraction of spurious detections.

For future versions of the \art\ source catalog, we plan to further adjust both the characteristic source size and the $S/N$ threshold, which define the optimal filter given by equation~(\ref{eq:art_phi_x}). They are currently equal to 53\arcsec\ and 4.5, respectively. We will use additional simulations of the source detection procedure in the \art\ all-sky survey for the adjustment.

\begin{figure*}
   \centering
    \includegraphics[width=0.98\textwidth]{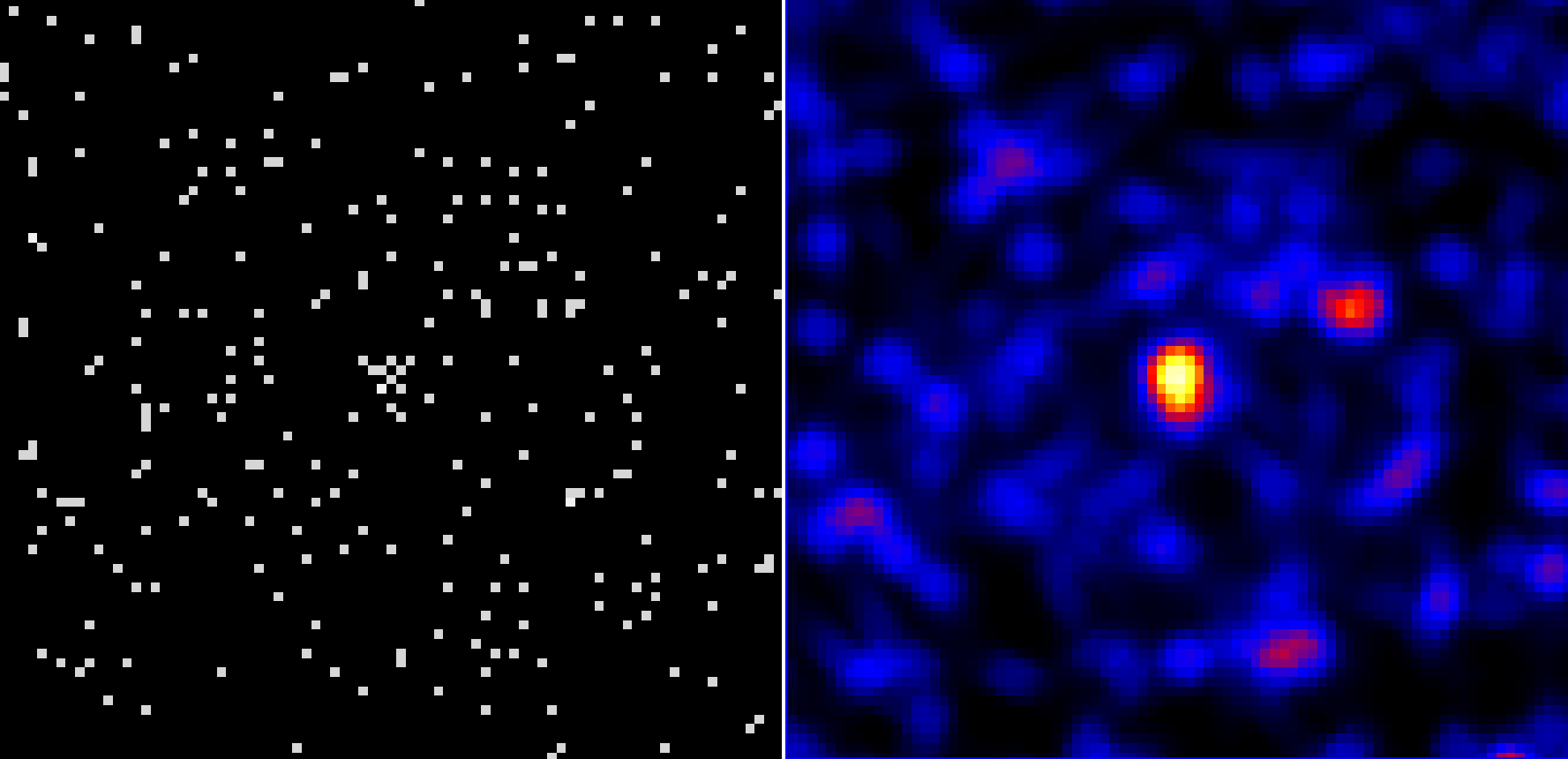}
    \caption{Example of a detection of a faint source using the Poisson optimal filter, described in \S\ref{s:detpoint}. Left: Raw photon image in the 4--12\,keV energy band. Right: Convolution with the optimal filter. The size of the images is approximately $20\arcmin\times 20\arcmin$.}
    \label{fig:img_faint}
\end{figure*}

\begin{figure}
   \centering
    \includegraphics[width=0.49\textwidth]{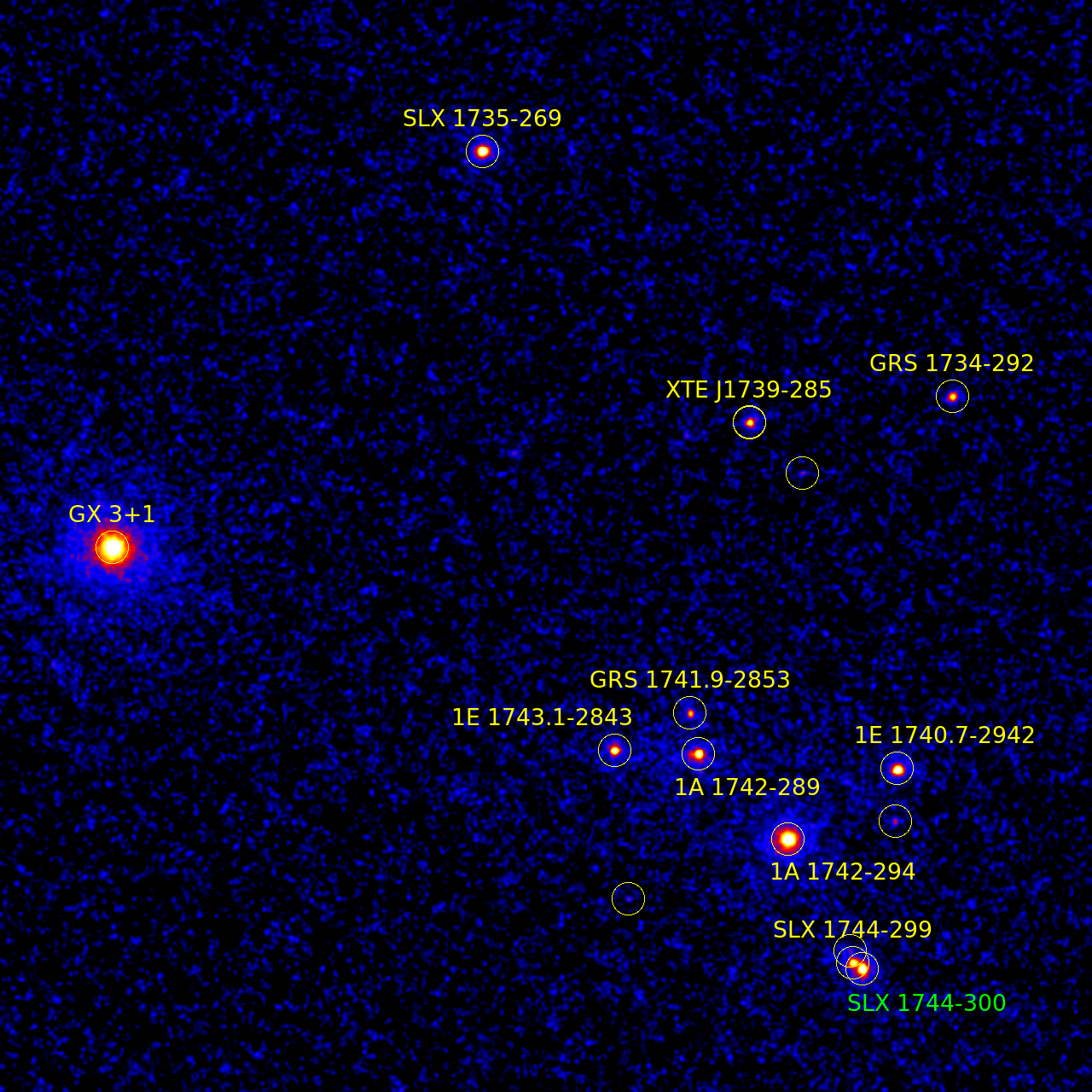}
    \caption{Map of the sky near the Galactic center obtained in ARTSS12 by convolution with the optimal filter in the 4--12\,keV energy band, in Galactic coordinates. The size of the image is approximately $4.5^\circ\times 4.5^\circ$. The detected sources are shown with circles, and the brightest sources ($S/N>10$) are additionally indicated by their common names.}
    \label{fig:img}
\end{figure}

Figure~\ref{fig:img_faint} shows an example of a detection of a faint source during the all-sky survey; the source with a flux of only a few photons is detected here with high significance. Figure~\ref{fig:img} shows a fragment of the resulting convolution image of the Galactic center region.

\subsection{Detection of extended sources}
\label{s:detext}

The filter given by equation~(\ref{eq:art_phi_x}) is only optimal for detection of point sources in the \art\ survey. Sources with an angular size larger than the \art\ PSF, that is, $\gtrsim 1\arcmin$, cannot be detected efficiently with this filter. Therefore we used the standard wavelet decomposition method (\texttt{wvdecomp}, \citealt{1998ApJ...502..558V}) to detect sources of large angular size (from $\sim 1$\,arcmin up to approximately half a degree).

Specifically, sources were found as peaks in the stacked decomposition images\footnote{A decomposition image is a wavelet convolution image with significant smaller size details subtracted and nonsignificant details of a given characteristic size masked with zeroes. See the \texttt{wvdecomp} documentation, available at an online repository archived at \href{https://doi.org/10.5281/zenodo.3610345}{doi:10.5281/zenodo.361034}} of larger than $4.5\sigma$ significance (Poisson statistics) at 6th to 8th wavelet scales, corresponding to angular scales approximately from 3\arcmin\ to 12\arcmin. The significance of a detection of extended sources was estimated using the noise images produced by \texttt{wvdecomp}.

We then correlated the candidate sources with catalogs of galaxy clusters, namely, the Meta-catalog of X-ray detected clusters of galaxies \citep[MCXC,][]{Piffaretti2011} and the Second Planck catalog of Sunyaev--Zeldovich sources \citep[PSZ2,][]{2016A&A...594A..27P}, as well as with the catalog of Galactic supernovae remnants (SNRs, \citealt{Green2019}). The adopted matching radius is 2\arcmin\ for MCXC, 5\arcmin\ for PSZ2, and the maximum size of a given remnant for the SNR catalog. In the last case, we also inspected the associations of X-ray sources with SNRs manually. The chosen matching radii take the typical positional uncertainties of object coordinates in the corresponding catalogs into account.

If an extended source was not found in any of these catalogs, we included it in the final \art\ source list only if its detection significance was higher than $5.5\sigma$. This threshold was chosen so that fewer than one spurious extended source is expected to be found in the survey, given the probability of false detection and the number of independent source detection cells (estimated assuming a minimum size of extended sources of 3\,arcmin) over the whole sky. Application of the second criterion led to the inclusion of two additional extended sources: the galaxy M31 (namely, unresolved X-ray emission from its central region), and the massive cluster of galaxies IGR\,J1744$-$3232. 

\subsection{Measurement of source X-ray fluxes}
\label{s:flux}

X-ray fluxes for detected point sources were measured in a circular aperture of 2\arcmin\ radius, which contains approximately 90\% of the total X-ray flux from the source at angular scales $<5\arcmin$. The flux was calculated from the count rate of background-subtracted photons in the aperture. As was explained in \S\ref{s:maps}, the vignetting effects were taken in account in the calculation of the exposure maps that are used to measure the source fluxes. The corresponding Poisson uncertainties were estimated using the approximate formulae from \cite{Gehrels86}. 

We calibrated the ratio of flux to count rate for the 4--12\,keV energy band using the available \art\ observations of the Crab nebula, taking the difference between the slopes of the spectrum of the Crab nebula ($\Gamma=2.1$) and the fiducial spectrum used in our source detection algorithm ($\Gamma=1.4$) into account. This resulted in a difference of a few percent in the fluxes. The Crab nebula is observed by \art\ as an extended source of $\approx 1\arcmin$ size, as expected. Its flux was computed from the spectral parameters given in \cite{2017ApJ...841...56M} and was fixed at this value.

The conversion factor of count rate to flux in general depends on the adopted procedure of selecting various types of \art\ detector events. For the criteria adopted in this work (see \S\ref{s:data}), a count rate of 1\,cnt~s$^{-1}$ corresponds to a flux of $3.34\times 10^{-11}$\,erg~cm$^{-2}$~s$^{-1}$ in the 4--12\,keV energy band.

The derived fluxes of some weak sources in the \art\ catalog are consistent with zero. These cases usually arise when the survey exposure varies significantly across the 2\arcmin\ aperture used for flux measurement (see above). We plan to improve our method of measuring source fluxes in future versions of the \art\ catalog.

The fluxes of extended X-ray sources were roughly estimated from the wavelet decomposition images. The large-scale part of the flux of an extended object that is missing in the decomposition image was estimated as the flux contained in the wings of a $\beta$-model \citep{1976A&A....49..137C} with $\beta=0.67$, 1\arcmin\ core radius, and the same flux in the central region (where the source is detected). The conversion of count rate to flux was made under the assumption of an optically thin $kT=5$\,keV plasma source spectrum. 

We compared the X-ray fluxes of galaxy clusters estimated in this way with the corresponding fluxes calculated (in the 4--12\,keV band) from the X-ray luminosities presented in the MCXC catalogue \citep{Piffaretti2011}, using the luminosity--gas temperature relation adopted from \cite{2009ApJ...692.1033V}. Our flux estimates prove to be accurate to better than 50\% for galaxy clusters. More accurate measurements of X-ray fluxes for extended \art\ sources must be made using an accurate model of the source image and spectrum. This is beyond the scope of this work.

\begin{figure*}
    \centering
    \includegraphics[width=\textwidth]{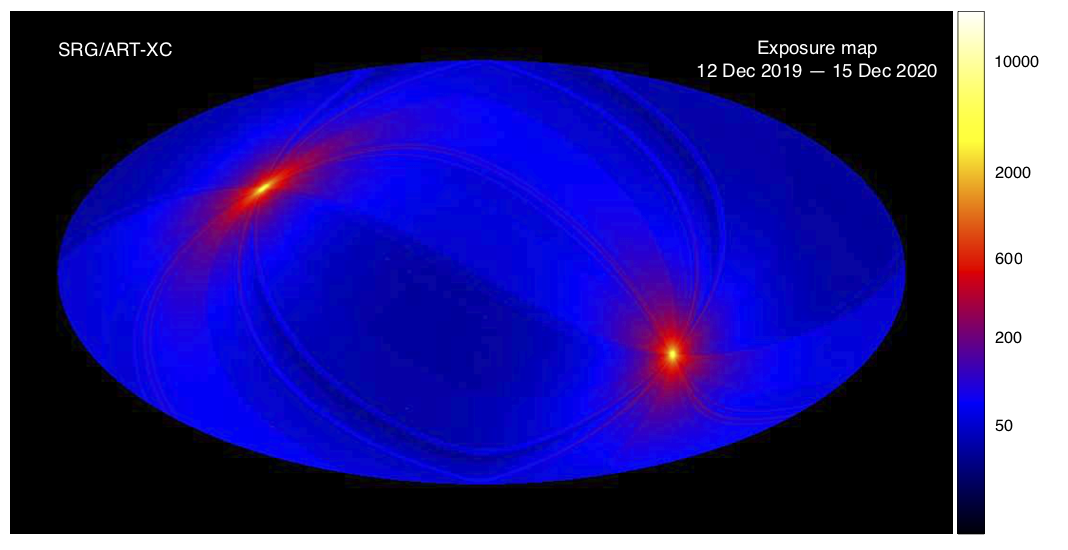}
    \caption{\art\ sky exposure map in Galactic coordinates in the 4--12\,keV energy band, vignetting corrected, after the first two all-sky surveys (ARTSS12). Exposure time is given in seconds (see the color scale on the right-hand side). Note the very high exposure accumulated near the ecliptic poles (bright spots) compared to the rest of the sky.}
    \label{fig:expmap}
\end{figure*}

\subsection{Merging X-ray source catalogs of individual sky tiles} 

Based on the procedures described above, we obtained catalogs of point and extended sources detected in each sky tile. These individual catalogs were then merged into a composite all-sky catalog, taking the overlap of tiles into account.

To this end, we first merged the catalogs of point and extended sources for each tile. If a point and an extended source are found at the same position in the sky (taking the source extent into account), we rejected the source with the lower X-ray flux. This allowed us to  eliminate false extended source detections associated with the broad PSF wings of bright point X-ray sources, as well as spurious point sources on top of bright extended sources.

The catalogs of sources detected in each tile were then merged into an all-sky catalog. To this end, sources that are located closer than 3\arcmin\ to the tile edge were rejected, and if a source was detected in two or more tiles, the detection at the larger distance from the tile edge was selected.

\subsection{Exposure map} 

The \srg\ observational strategy in sky survey mode is described in \cite{Sunyaev21}. It results in a relatively uniform exposure near the ecliptic plane, which varies from $\approx30$ to $\approx80$\,s with a median of $\approx 60$\,s (vignetting corrected) after one year of the \art\ survey. In the regions near the ecliptic poles, the exposure is much higher, reaching $\sim 17$\,ks, and it is nonuniform. 

\begin{figure*}
    \centering
    \includegraphics[width=0.46\textwidth]{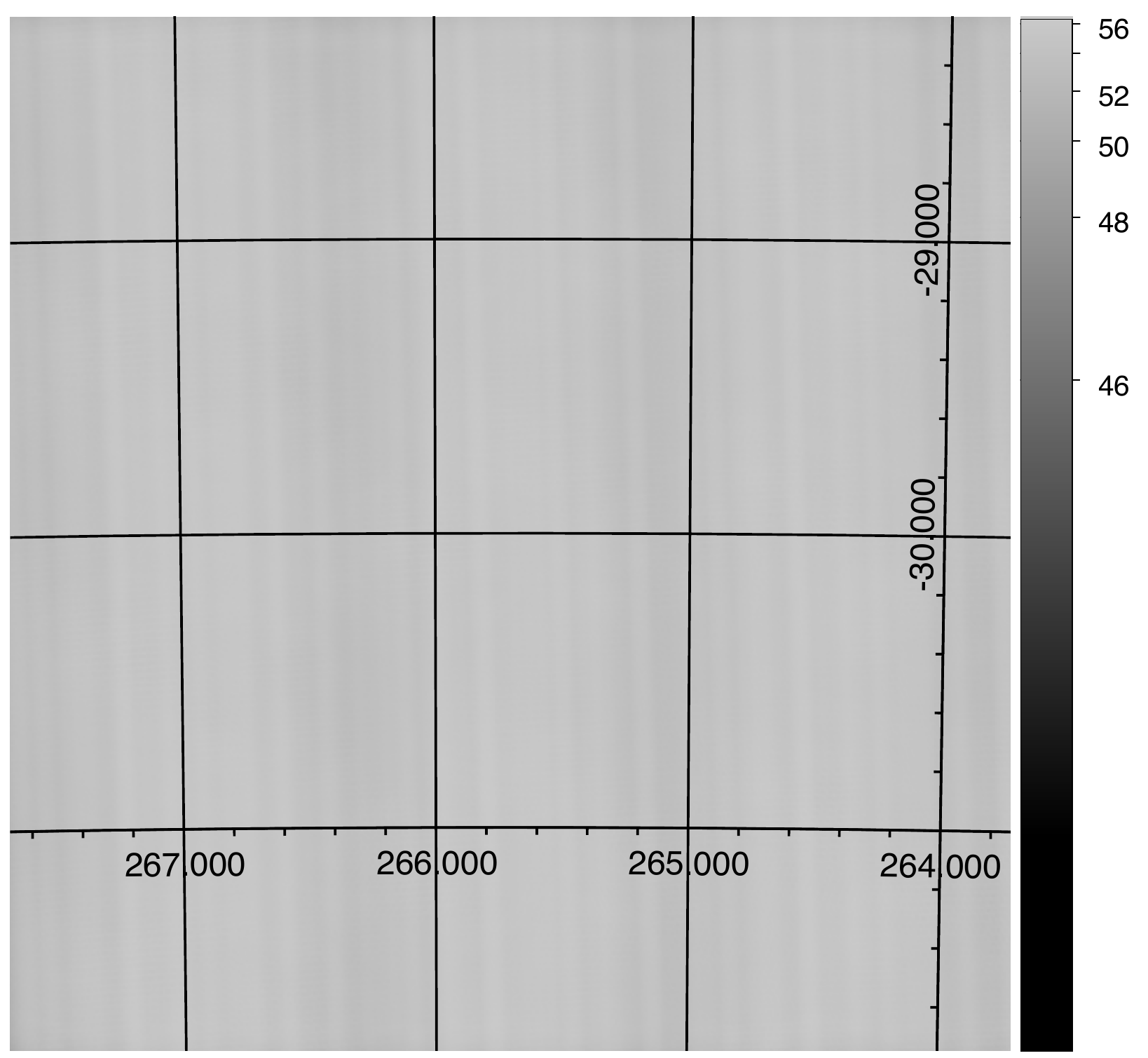}
    \includegraphics[width=0.50\textwidth]{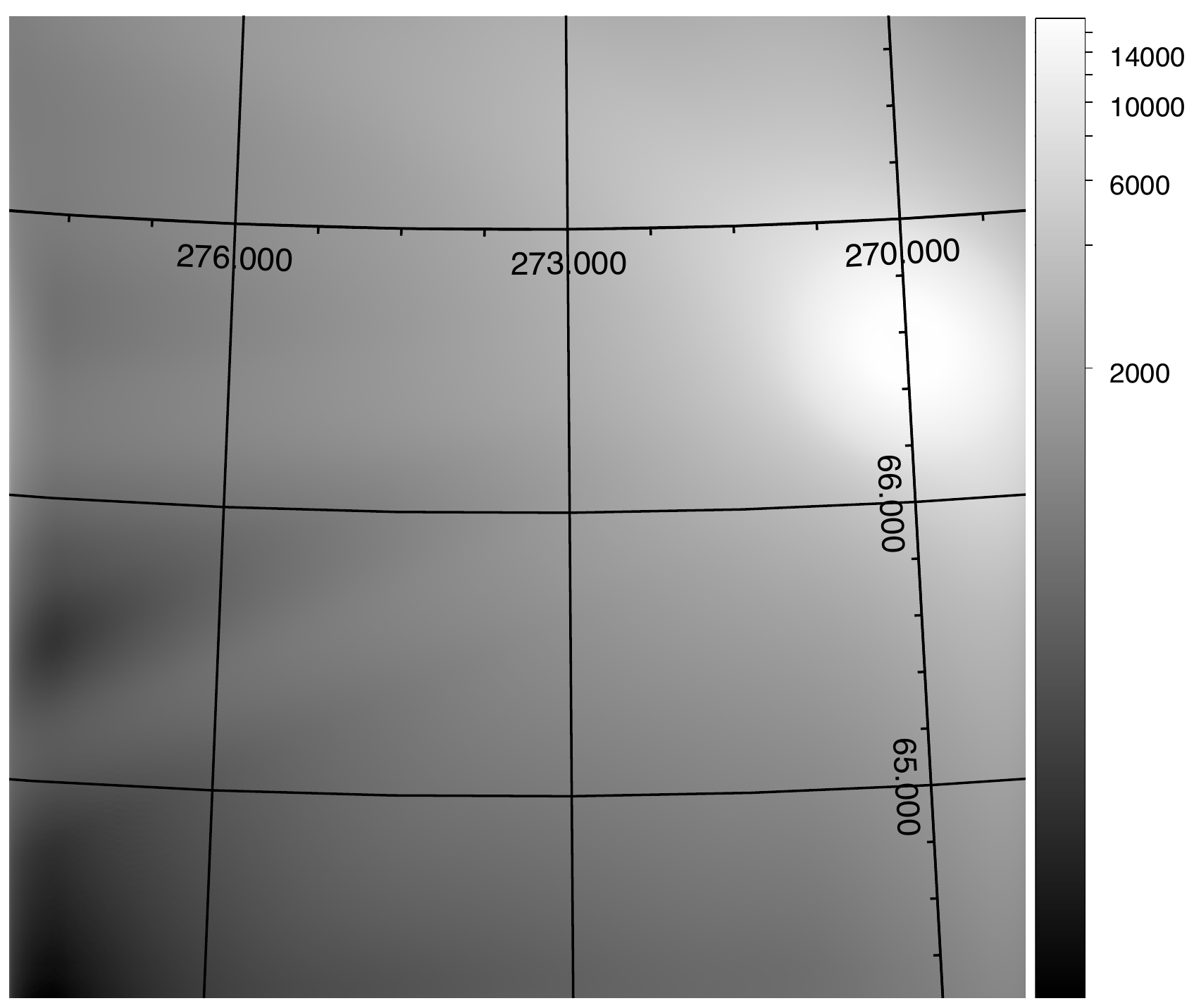}
    \caption{First-year survey $3^\circ\!.6\times 3^\circ\!.6$ exposure maps near the Galactic center (left) and near the NEP (right), calculated for the 4--12\,keV energy band, vignetting corrected. Exposure time is given in seconds (see the color scales next to the panels). The equatorial coordinate system grid is shown.}
    \label{fig:exp}
\end{figure*}

Figure~\ref{fig:expmap} shows the all-sky vignetting-corrected exposure map in the 4--12\,keV band for the first year of the survey. Zoomed-in fragments of this map near the Galactic center and near the north ecliptic pole (NEP) are shown in Fig.~\ref{fig:exp}.

\section{Construction of the final catalog}

\subsection{Source detection thresholds}
\label{s:thresh}

\begin{figure}
    \centering
    \includegraphics[width=0.49\textwidth]{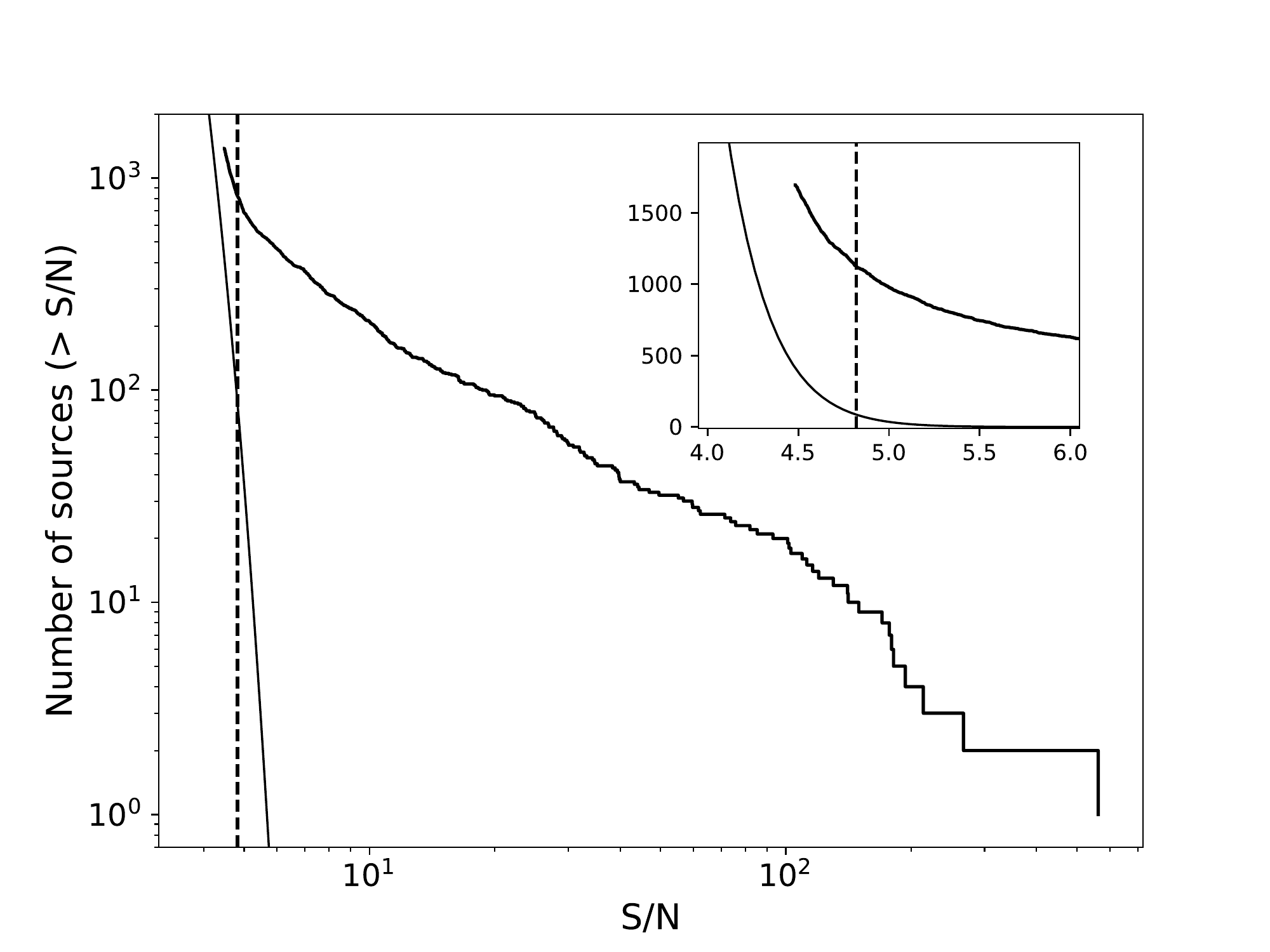}
    \caption{
    Number of point sources as a function of detection significance in ARTSS12 in the 4--12\,keV band (thick solid line). The dashed vertical line indicates the adopted threshold of $S/N=4.82$, which corresponds to the expected 10\% fraction of spurious sources. The thin solid line shows the expected number of spurious sources as a function of the detection significance. The inset shows a zoom-in near the detection threshold.}
    \label{fig:sig_nsrc}
\end{figure}

The size of the resulting catalog depends on the adopted threshold for source detection. We originally detected point sources down to the $S/N=4.5$ significance limit, with $S/N$ defined in \S\ref{s:detpoint}. At this level, a large number of spurious sources are expected to appear on the \art\ all-sky map. By raising the threshold, we can improve the purity of the catalog at the cost of reducing its size.  

Figure~\ref{fig:sig_nsrc} shows the total number of point sources detected by \art\ in ARTSS12 in the 4--12\,keV band as a function of the significance limit. The total number reaches 1335 for $S/N > 4.5$ when the expected fraction of false detections (thin solid curve in Fig.~\ref{fig:sig_nsrc}) is as high as $\approx26.6\%$. This fraction was estimated from the results of extensive Monte Carlo simulations of empty fields (\S\ref{s:detpoint}). We thus decided to raise the detection limit to $S/N=4.82$, so that the expected fraction of spurious sources is 10\%. At this threshold, the total number of point sources detected in ARTSS12 in the 4--12\,keV energy band is 821 (with $\approx 241$ real sources expected to be lost compared to the $S/N=4.5$ threshold). For comparison, a still cleaner sample with 1\% of spurious sources consists of 571 point sources with $S/N > 5.34$.

It is difficult to specify a similarly strict statistical threshold for the inclusion of extended sources into the \art\ catalog. Based on the algorithm described in \S\ref{s:detext}, a total of 46 extended sources detected in ARTSS12 were included in the catalog. All of these are previously known astrophysical objects and have been detected in X-rays before. 

\subsection{Identification and classification of \art\ sources}
\label{s:ident}

The \art\ localization accuracy is better than 40\,arcseconds for the weakest detected sources (see \S\ref{s:loc}). This enables a fairly straightforward search for likely counterparts of \art\ sources in other wavebands. 

To this end, we cross-correlated the \art\ source catalog with the SIMBAD Astronomical Database \citep{simbad} and NASA/IPAC Extragalactic Database\footnote{The NASA/IPAC Extragalactic Database is funded by the National Aeronautics and Space Administration and operated by the California Institute of Technology.} (NED) as well as with the X-ray astronomy database provided by the High Energy Astrophysics Science Archive Research Center (HEASARC). The search for counterparts was carried out within 40\arcsec\ from the positions of \art\ sources. The majority of the \art\ sources could be unequivocally associated with a single previously known X-ray source within this matching radius. If no known X-ray sources or more than one potential X-ray counterparts from external catalogs were found within the matching radius, we searched the literature in an attempt to identify and classify the \art\ source. For the same purpose, we also searched (within 40\arcsec) for possible counterparts in other wavebands in catalogs of optical, infrared, and radio all-sky surveys, namely Gaia Early Data Release 3 (Gaia EDR3; \citealt{Gaia2021}), AllWISE \citep{Cutri2014}, and the Faint Images of the Radio Sky at Twenty-cm survey (FIRST, \citealt{Helfand2015}), the NRAO VLA Sky Survey (NVSS, \citealt{Condon1998}), and the Sydney University Molonglo Sky Survey (SUMSS, \citealt{Mauch2003}). 

The resulting identifications and classifications for \art\ sources as well as redshifts for extragalactic sources were adopted from SIMBAD and/or NED. In the case of dubious identification or classification, for recently discovered X-ray sources, and for newly discovered \art\ sources, we used additional information found in the literature and/or inferred from our multiwavelength cross-correlation analysis described above. All these cases are discussed in detail in Appendix~\ref{s:notes}, where the corresponding references are also provided.

\subsection{Follow-up campaign}

We have set up a program of optical follow-up observations for new X-ray sources discovered by \art\ and for the \art\ sources that were known as X-ray sources from previous missions, but whose nature remained unknown or uncertain. 
In the northern part of the sky, this program is being carried out at two telescopes that compose the ground segment of the \srg\ mission: the Sayan observatory 1.6 m telescope (AZT-33IK, \citealt{Burenin16}), operated by the Institute of Solar-Terrestrial Physics of the Siberian branch of the Russian Academy of Sciences, and the Russian-Turkish 1.5 m telescope (RTT-150), operated jointly by the Kazan Federal University, the Space Research Institute (IKI, Moscow), and the TUBITAK National Observatory (TUG, Turkey). 

The first results of this program, including a number of newly identified AGN and CVs, have been reported by  \citealt{2021AstL...47...71Z,2021arXiv210705611Z}; Zaznobin et al., in preparation. This information has been included in the \art\ source catalog and in the corresponding notes to the catalog.

\begin{figure*}
    \centering
    \includegraphics[width=\textwidth]{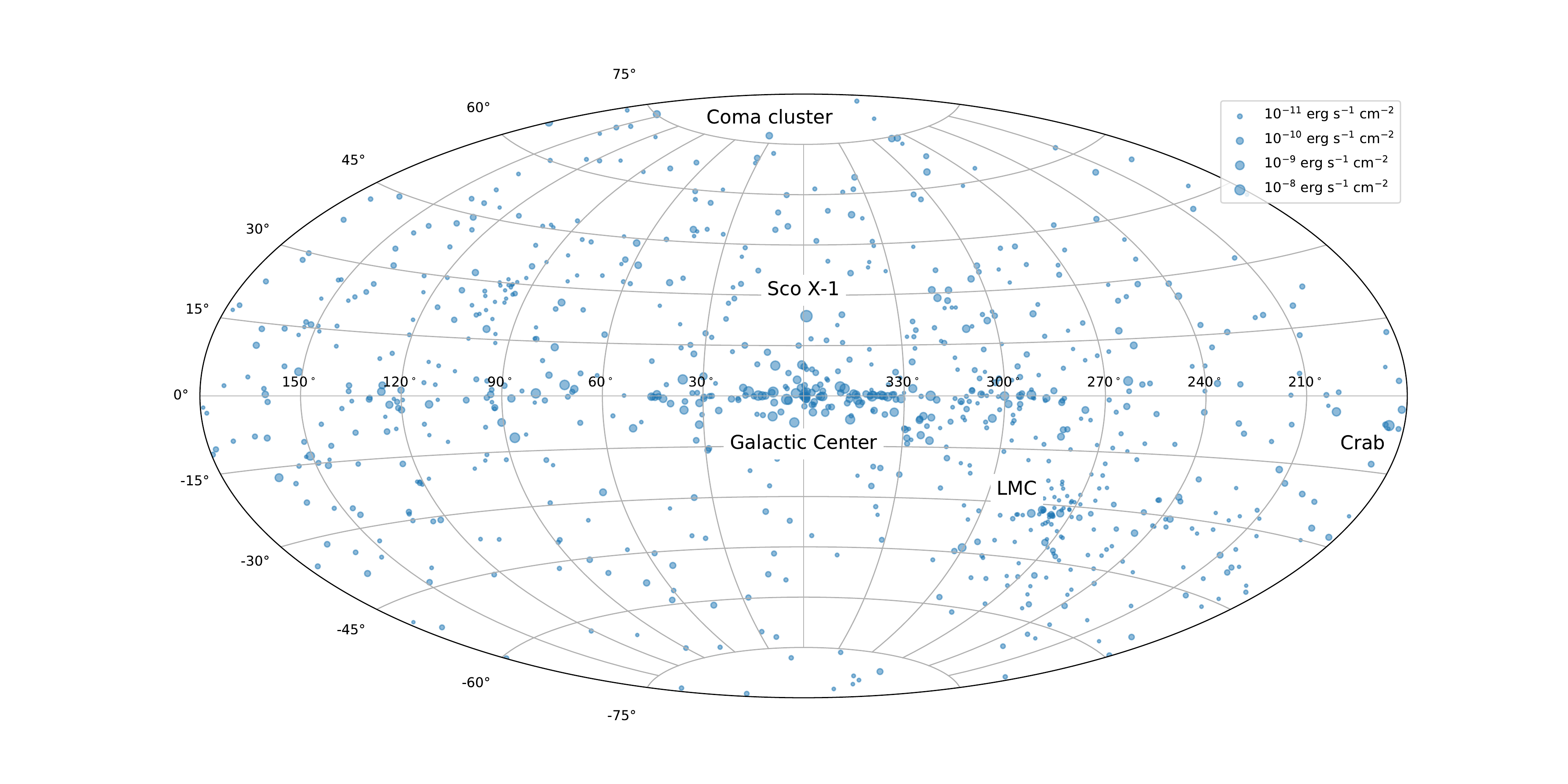}
    \caption{Positions (in Galactic coordinates) of the X-ray sources detected by \art\ in the 4--12\,keV energy band during the first year of the all-sky survey. The symbol size reflects the X-ray brightness of a source, as indicated in the legend in the top right corner.}.
    \label{fig:map}
\end{figure*}

\section{Source catalog}
\label{s:catalog}

The content of the catalog is as follows:

{\it Column (1) ``Id''} : The source sequence number in the catalog.

{\it Column (2) ``Name''} : The source name in the \art\ catalog (prefix ``SRGA'' followed by the source coordinates). New X-ray sources, that is, sources detected for the first time in X-rays by \srg/\art, are highlighted in bold text.

{\it Columns (3,4) ``RA, Dec''} : The equatorial (J2000) coordinates of the source.

{\it Columns (5) ``S/N''} : The detection significance of the source.

{\it Column (6) ``Flux''} : The time-averaged source flux in the 4--12\,keV energy band and the corresponding lower and upper uncertainties (in parentheses). For some sources, only an upper limit on the (aperture) flux is reliably obtained (the cases for which the quoted lower error is equal to the flux), even though the source has surpassed the detection significance threshold (see \S\ref{s:flux}).

{\it Column (7) ``Common name''} : The common name of the source, if available. 

{\it Column (8) ``Redshift''} : The source redshift (for extragalactic objects), if known. 

{\it Column (9) ``Class''} : The general astrophysical class of the object: LMXB (HMXB), that is, low- (high-) mass X-ray binary; X-RAY BINARY, that is, X-ray binary of uncertain type; CV, that is, cataclysmic variable or symbiotic binary; SNR, that is, supernova remnant; SNR/Pulsar, that is, supernova remnant with a central pulsar (when both may contribute to the X-ray emission); MAGNETAR, that is, a  magnetar (anomalous X-ray pulsars and soft gamma-ray repeaters); STAR, that is, an active star (of various types, excluding the previously listed types of stellar objects); SFR, that is,  a star-forming region; SEYFERT, that is, AGN of the Seyfert or (rarely) LINER type; AGN, that is, unclassified AGN; BLAZAR, that is, a  beamed AGN (BL Lac or flat-spectrum radio quasar); and CLUSTER, that is, a  cluster of galaxies. UNIDENT lists an unclassified source. A question mark indicates that the quoted classification is tentative; see comments on all these cases in Appendix~\ref{s:notes}.

In addition to this key information, we plan to maintain an extended online version of the \art\ catalog at \url{http://srg.cosmos.ru} where additional information will be presented. Specifically, we plan to provide cross-matches with external catalogs in X-ray and other wavebands for each ART-XC source as well as \art\ maps of the corresponding region of the sky.
Figure~\ref{fig:map} shows the celestial distribution of the \art\ sources detected during the first year of the all-sky survey. 

\subsection{Source number--flux function and the sensitivity of the survey}
\label{s:counts}

\begin{figure}
    \centering
    \includegraphics[width=0.49\textwidth]{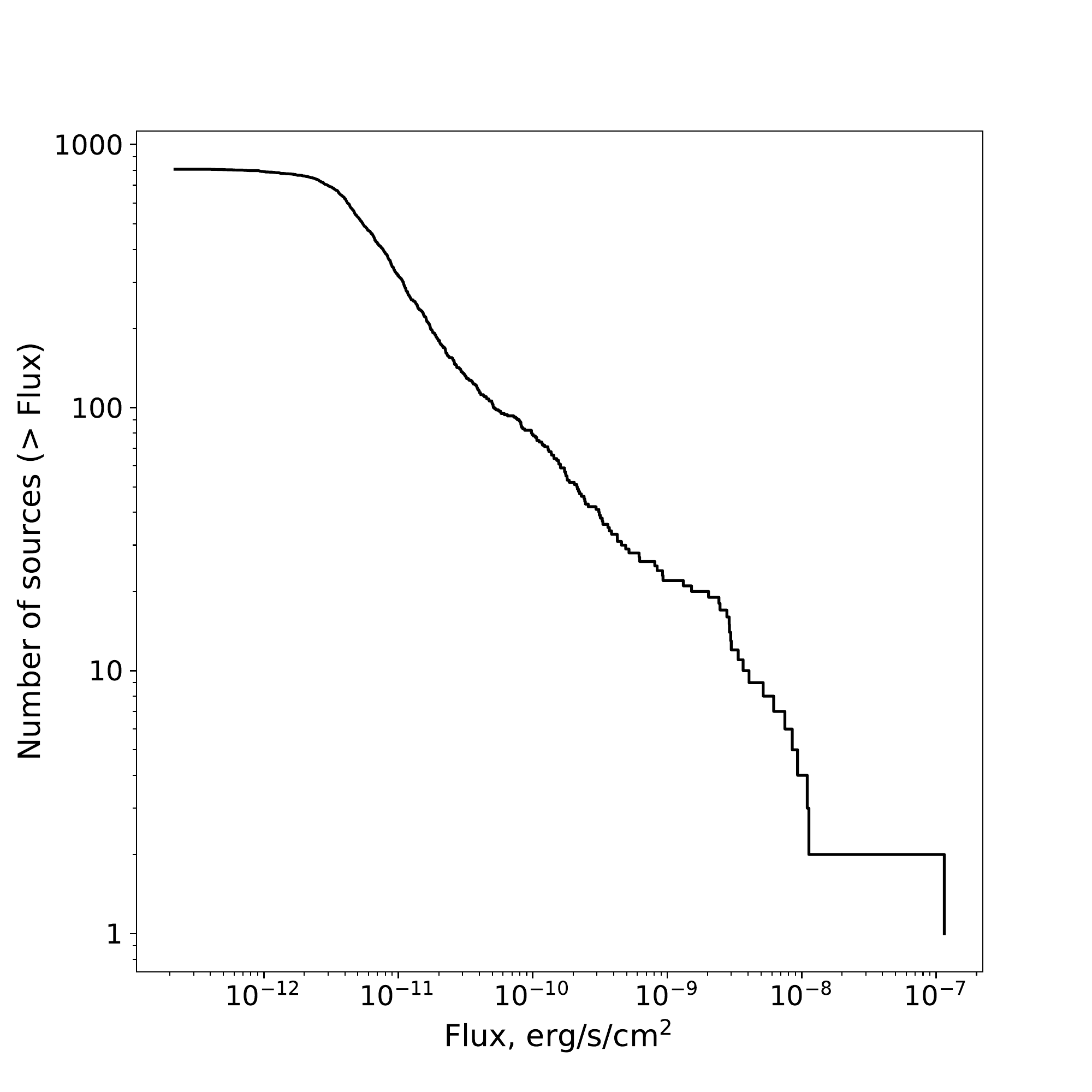}
    \includegraphics[width=0.49\textwidth]{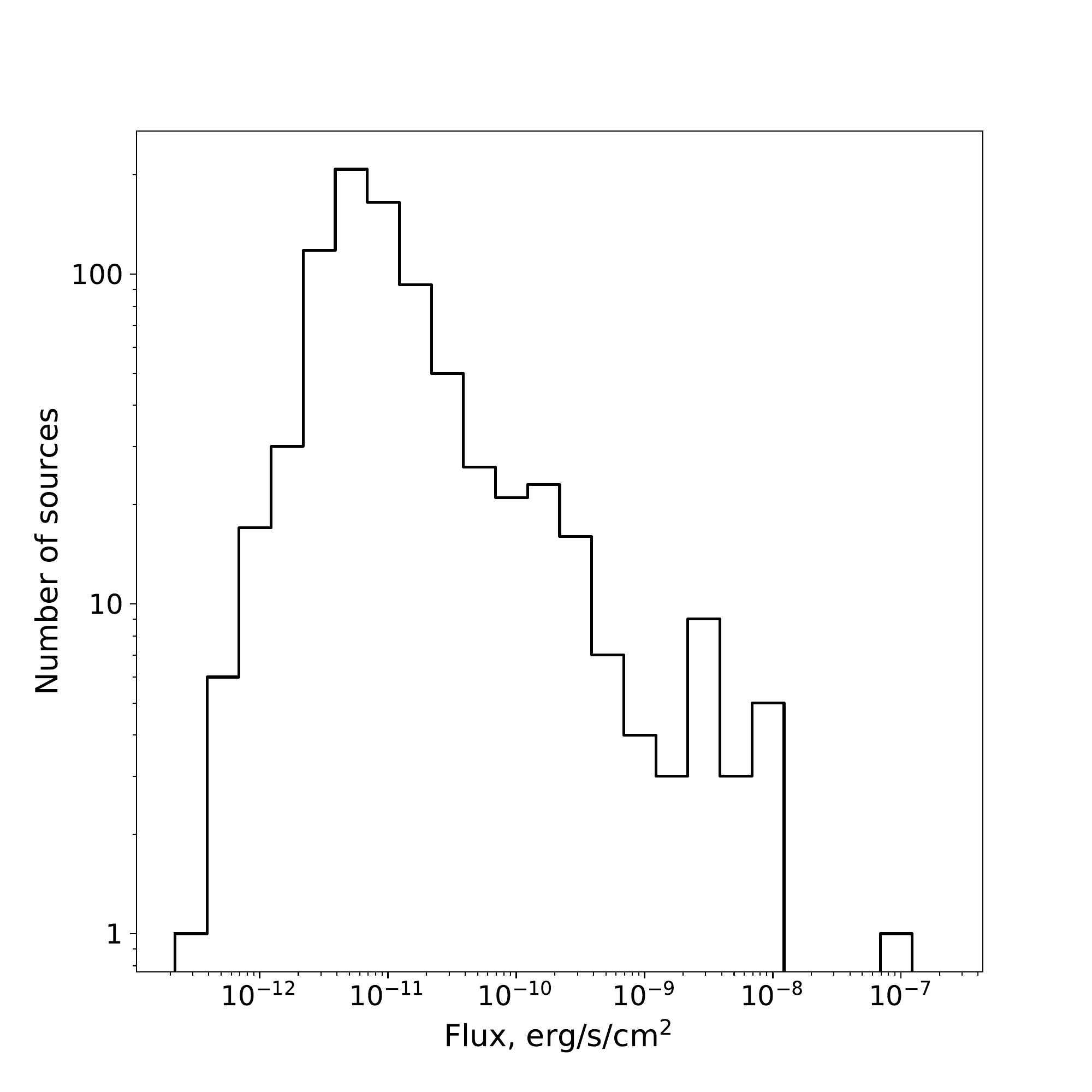}
    \caption{Cumulative (top) and differential (bottom) flux distributions of the point sources detected in ARTSS12 in the 4--12\,keV  energy band. Sources with only upper limits on the flux were excluded. The differential distribution peaks at a flux of $(4.1\pm0.5)\times10^{-12}$\,\flux.}
    \label{fig:flux_nsrc}
\end{figure}

Figure~\ref{fig:flux_nsrc} shows the cumulative and differential flux distributions of the point sources detected in ARTSS12 in the 4--12\,keV energy band. The median flux of point sources (above the adopted threshold of $S/N=4.82$) is $7.6 \times 10^{-12}$\,\flux. 

\begin{figure}
    \centering
    \includegraphics[width=0.5\textwidth]{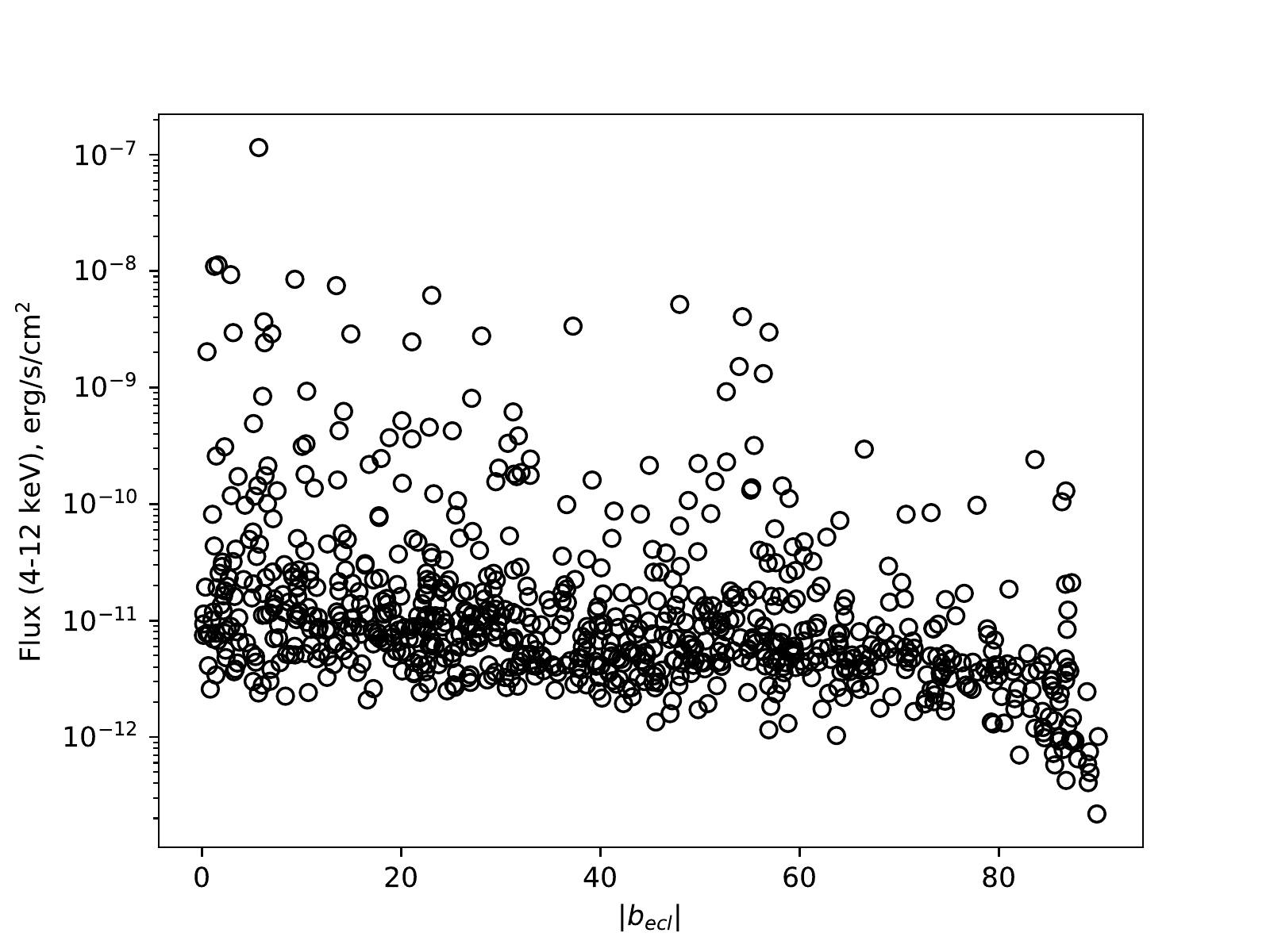}
    \caption{Distribution of the point sources detected in ARTSS12 in the 4--12\,keV energy band over the flux in this band, and the ecliptic latitude. Sources with only upper limits on the flux were excluded.}
    \label{fig:b_flux}
\end{figure}

Figure~\ref{fig:b_flux} shows the distribution of the point sources on the plane of ecliptic latitude -- X-ray flux. It is evident that the sensitivity of the survey monotonically increases from $\sim 4\times 10^{-12}$\,\flux\ near the ecliptic plane (at $|b_{\rm ecl}|<30^\circ$) to $\sim 8\times 10^{-13}$\,\flux\ near the ecliptic poles (at $|b_{\rm ecl}|>80^\circ$). The quoted values are the median fluxes of the sources (excluding the few for which only upper limits on the flux are available) detected with $4.82<S/N<5$, that is, near the adopted threshold. More accurate estimates of the sky coverage as a function of sensitivity for the \art\ survey will be made through extensive Monte Carlo simulations of the source detection procedure and will be presented elsewhere. 

\subsection{Source classes}

\begin{table}
\caption{Statistics of the sources listed in the combined catalog.}
\label{tab:source_types}
\centering

\begin{tabular}{cl}
\hline
Type & Count (Notes) \\
\hline
LMXB &  89 \\
HMXB & 72 \\
X-ray binary (unclassified) &  6\\
CV & 102 \\
Star &  34 \\
Magnetar & 5 \\
SNR, SNR/Pulsar & 15 \\
Star-forming region & 3 \\
Galaxy & 1 (M31) \\
ULX & 1 (M82 X-1) \\
Seyfert galaxy & 243 \\
AGN (unclassified) & 40 \\
Blazar & 87 \\
Galaxy cluster & 52 \\
Unidentified & 117 \\
\hline
\end{tabular}
\end{table}

\begin{table}
\caption{Statistics of sources by category.}
\label{tab:source_category}
\centering

\begin{tabular}{cl}
\hline
Category & Count \\
\hline
Galactic & 313 \\
Local Group & 18 \\
Extragalactic & 419 \\
\hline
\end{tabular}
\end{table}

Table~\ref{tab:source_types} summarizes the statistics of objects of various classes present in the \art\ catalog. One hundred seventeen sources, or 13\%, remain unidentified or unclassified so far, and many of these may be spurious. Most of the remaining 750 sources, namely 56\%, have an extragalactic nature (including objects located in the Local Group of galaxies, see Table~\ref{tab:source_category}), but the number of Galactic objects is not much lower. The largest groups within the Galactic category are CVs (and symbiotic binaries), LMXBs, and HMXBs, although there are also a significant number of hot stars and supernova remnants/pulsars. The extragalactic objects are dominated by AGN, including relativistically beamed ones (i.e., blazars). The second largest group consists of clusters of galaxies. 

\begin{figure}
    \centering
    \includegraphics[width=0.5\textwidth]{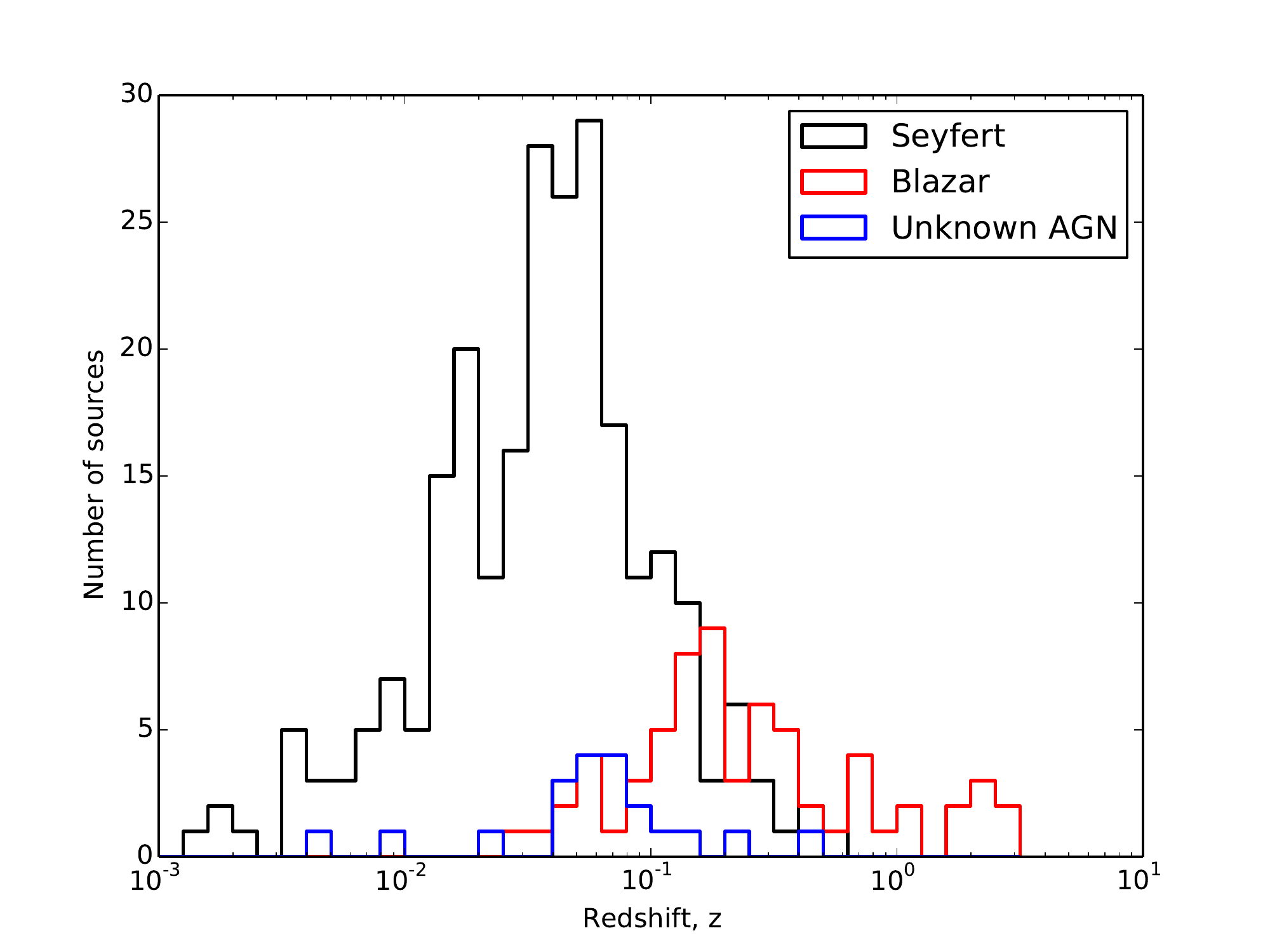}
    \caption{Redshift distribution of the AGN detected by \art: Seyfert galaxies (black), blazars (red), and unclassified AGN (blue).}
    \label{fig:agn_z}
\end{figure}

Figure~\ref{fig:agn_z} shows the redshift distribution of the AGN. The median redshift of the unbeamed (i.e., Seyfert-like) AGN is 0.04, and it is 0.19 for the blazars. This shows that \art\ mainly probes the AGN population in the nearby Universe. Redshifts are still missing for 40 AGN and AGN candidates in the \art\ catalog.

\subsection{Localization accuracy}
\label{s:loc}

\begin{figure}
    \centering
    \includegraphics[width=0.5\textwidth]{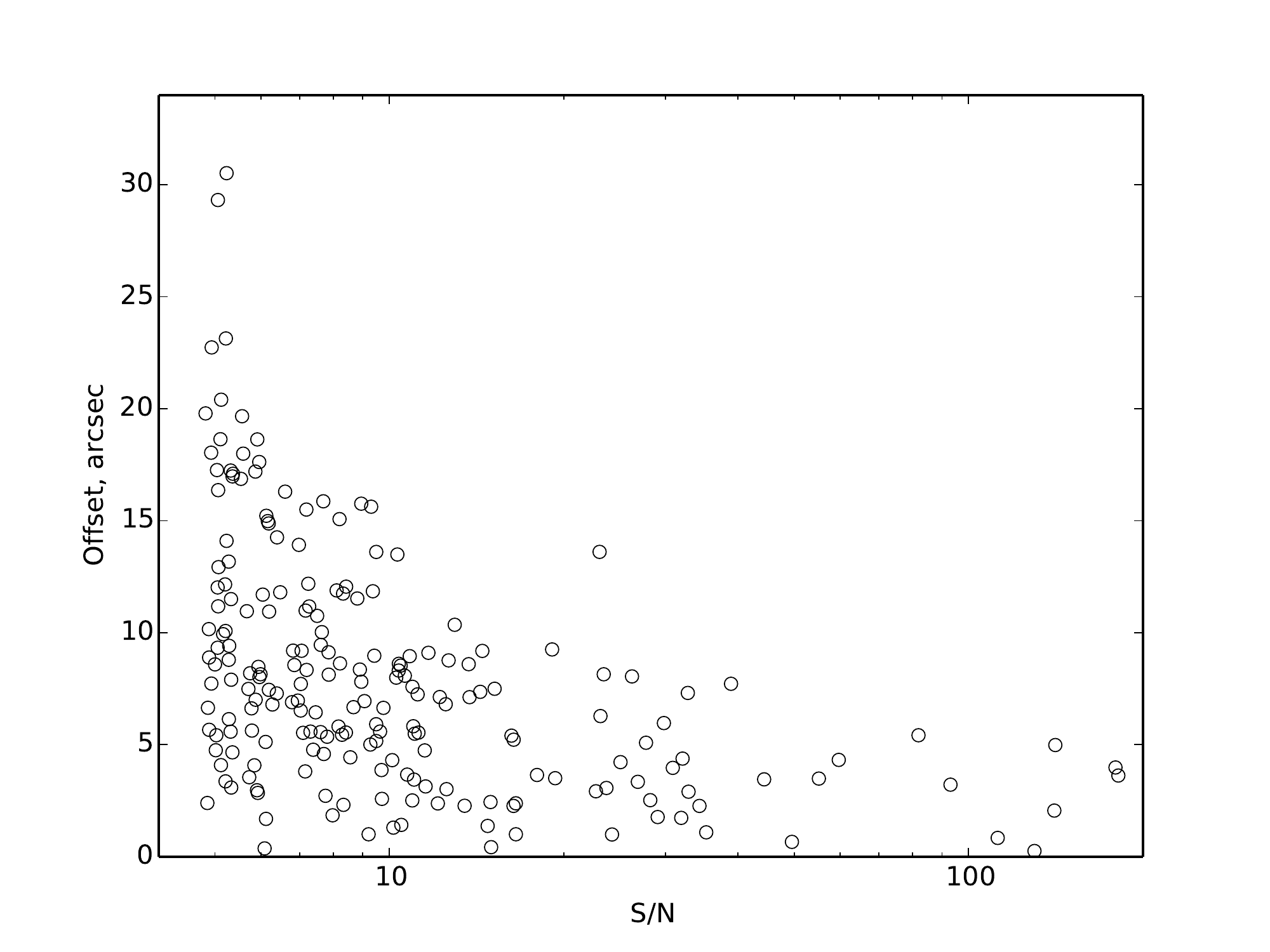}
    \caption{X-ray--optical offsets as a function of source detection significance for \art\ sources with point-like Galactic counterparts.}
    \label{fig:offset}
\end{figure}

To evaluate the localization accuracy of \art\ during the all-sky survey, we selected those sources from the \art\ catalog that are associated with a point-like Galactic object (i.e., a star, CV, X-ray binary, etc.) and whose optical counterpart (found in SIMBAD or in the literature cited in Appendix~\ref{s:notes}) is present in the Gaia DR3 catalog (i.e., has a visual magnitude $G\la 21$). We then compared the \art\ positions of these X-ray sources with the Gaia positions of their optical counterparts. In total, 209 objects were included in this test.

Figure~\ref{fig:offset} shows the resulting X-ray--optical positional offsets as a function of source detection significance. The typical accuracy is $\leq 15 \arcsec$ for $S/N> 6$ and it does not exceed $35 \arcsec$ for the weakest sources. Based on this, we adopt 40\arcsec\ as a minimum search radius for cross-matching with external source catalogs.

\subsection{Cross-match with other X-ray and gamma-ray catalogs}

The majority of sources in the \art\ catalog were already known from previous X-ray missions. To evaluate this quantitatively, we cross-correlated the 821 point sources from the \art\ catalog with a number of all-sky or nearly all-sky X-ray and gamma-ray surveys, namely the Second \rosat\ all-sky survey (2RXS; \citealt{Boller16}), the \xmm\ slew survey (XMMSL2; \citealt{Saxton08}, table XMMSLEWCLN in HEASARC), the combined MAXI/GSC 7-year all-sky source catalog (3MAXI; \citealt{Kawamuro18,Hori18}, table MAXIGSC7YR in HEASARC), the 105-month \swift/BAT all-sky survey (hereafter, Swift105mo; \citealt{Oh18}), the \integral/IBIS 17-year all-sky survey (hereafter, INTEGRAL17yr; Krivonos et al., in preparation), and the \fermi/LAT 10-year all-sky survey (4FGL-DR2; \citealt{2020ApJS..247...33A,2020arXiv200511208B}). We excluded the extended \art\ sources from the cross-matching analysis to avoid complications related to the uncertainty in their size and morphology. All these 46 extended objects are well-known X-ray sources (see \S\ref{s:detext}).

We regarded the outcome of cross-matching a given \art\ source with a given external catalog as positive if at least one source from that catalog was found within 40\arcsec\ from the \art\ position or if the \art\ source proved to lie within the $90\%$ localization region of a source from the external catalog. Hence, the adopted cross-matching algorithm can be described as follows: if $S<40''$~OR~$S<E_{90}$, where $S$ is the angular separation between an \art\ source and a candidate counterpart, and $E_{90}$ is the 90\% position error of the counterpart. The first of these conditions mainly pertains to 2RXS and XMMSL2 because the position errors of the sources in these catalogs are comparable to or better than those of \art\ sources (\S\ref{s:loc}), while the second criterion mainly governs the cross-matching with the other catalogs mentioned above, which are characterized by larger positional uncertainties.

For a number of 3MAXI sources at a low Galactic latitude without a reported position error in \cite{Hori18}, we estimated $E_{90}$ as 114\arcmin/$S_{\rm det,\,4-10}$, where $S_{\rm det,\,4-10}$ is the \maxi/GSC source detection significance in the 4--10\,keV band. We found that this dependence describes the correlation between these two quantities for 3MAXI sources well. The position error radii for Swift105mo were estimated from eq.~(1) in \cite{Oh18}, with an additional factor of 1.42 to convert from 1$\sigma$ into 90\% confidence. The positional uncertainties for INTEGRAL17yr were estimated in different ranges of source detection significance according to \cite{Krivonos2007}. For simplicity, we converted the ellipsoidal position errors provided in the 4FGL-DR2 catalog into effective error radii corresponding to the same area of the localization region and excluded extended 4FGL-DR2 sources from the cross-matching analysis.

There are no cases of multiple cross-matches (i.e., more than one counterpart for a given \art\ source) with the 2RXS, 3MAXI, Swift105mo, INTEGRAL17yr, and 4FGL-DR2 catalogs. However, two \art\ sources SRGA\,J174726.0--300002 and SRGA\,J174726.2--300245 (associated with SLX\,1744--299 and SLX\,1744--300, respectively) are blended into one \integral\ source (we counted this case as 2 cross-matches). Additionally, there are 36 multiple cross-matches with XMMSL2 when at least two \xmm\ sources are found within the search region around a given \art\ source (each such group is regarded as one cross-match). The vast majority of these cases are evidently caused by the imperfect merging of individual \xmm\ slew detections into ``unique sources'' in the XMMSL2 catalog.

Table~\ref{tab:cross_match} provides the results of our cross-matching analysis. The largest overlap of the \art\ survey is observed with the 2RXS, Swift105mom, and XMMSL2 catalogs. The result of cross-matching with the 4FGL-DR2 catalog of gamma-ray sources is particularly interesting. Eighty-seven \art\ sources associated with \fermi\ sources have known identifications, including 16 Galactic objects (3 SNRs/pulsars, 8 LMXBs, 3 HMXBs, the star eta Car, and the \art\ source SRGA\,J181637.1--391248, which is probably associated with variable star V371 CrA at 414\arcsec\ from 4FGL\,J1816.1--3908), 68 extragalactic objects (9 Seyfert galaxies, 58 blazars, and one ULX -- M82 X-1), and 3 SNRs/pulsars in the Large Magellanic Cloud (LMC; SNR\,J052501--693842, PSR\,J0537--6910, and PSR\,B0540--69). The \art\ catalog includes 87 blazars and blazar candidates, and the majority of them (58) have a \fermi\ gamma-ray counterpart. Additionally, one unidentified \art\ source, SRGA\,J101132.4--442258, is spatially consistent with 4FGL\,J1011.1--4420.

The vast majority of the cross-matches are expected to be real. To estimate the number of accidental matches, we can make the crude assumption that the sources in all of the catalogs under consideration (including the \art\ catalog) are distributed uniformly over the sky. Then the expected number of spurious matches can be found as the area of the search region that was used for cross-correlation divided by the area of the sky and multiplied by the number of sources in the \art\ catalog and by the number of sources in the matching catalog. As mentioned before, the search radius is typically 40\arcsec\ for 2RXS and XMMSL2. For the other catalogs, we can use the median radius of the 90\% localization regions, which is equal to 613\arcsec, 343\arcsec, 179\arcsec, and 192\arcsec\ for 3MAXI, Swift105mo, INTEGRAL17yr, and 4FGL-DR2, respectively. In this way, we predict that the number of spurious matches with \art\ point sources is $\sim 1.0$, $\sim 0.2$, $\sim 1.6$, $\sim 0.9$, $\sim 0.1$, and $\sim 1.0$ for 2RXS, XMMSL2, 3MAXI, Swift105mo, INTEGRAL17yr, and 4FGL-DR2, respectively, which is lower than 2 in each case.

However, the assumption of a uniform distribution of X-ray sources over the sky might be too crude because the Galactic sources are concentrated around the Galactic plane and the Galactic center (see Fig.~\ref{fig:map}) and because an additional several dozen \art\ sources are concentrated around the ecliptic poles (see \S\ref{s:ecl} below) because these regions receive a deeper exposure in the \srg\ survey. We can try to take this spatial nonuniformity into account by shifting the positions of the \art\ point sources by some small amount and repeating our cross-matching analysis. Specifically, we shifted the \art\ positions in random directions by 45\arcmin \ because this angular distance, on the one hand, is larger than the median 90\% position uncertainty of the 3MAXI catalog by a factor of $\approx 4.4$ and by larger factors for the other catalogs considered here, and on the other hand, it is small compared to the effective thickness of the distribution of X-ray sources near the Galactic plane and the effective size of the \art\ deep fields near the ecliptic poles. Using this alternative method, we found somewhat larger numbers of spurious cross-matches of the \art\ catalog with the external catalogs, namely 2, 0, 3, 1, 0, and 4 for 2RXS, XMMSL2, 3MAXI, Swift105mo, INTEGRAL17yr, and 4FGL-DR2, respectively. Although in most of these cases the position error of the matching source is much smaller than its angular distance from the original (unshifted) position of the \art\ source, some of the 3MAXI and 4FGL-DR2 matches have position errors $\sim 1/3$--1/2 of that distance and thus might be real associations with the \art\ sources. Therefore the derived numbers for 3MAXI and 4FGL-DR2 should be considered upper limits. Taking this into account, we provide in Table~\ref{tab:cross_match} our final estimates of the number of random coincidences of \art\ sources with sources from the external catalogs.

One hundred forty-one \art\ point sources have not been detected in any of the six all-sky surveys listed in Table~\ref{tab:cross_match}. In this context, we recall that $\sim 80$ of these are expected to be spurious sources.

We finally note that all of the surveys listed in Table~\ref{tab:cross_match}, except for 2RXS, were carried out over at least 8 years. Some of them are characterized by a very low duty cycle. In particular, XMMSL2 spans 13\,years, and the net exposure time was just a few months. For comparison, the \art\ catalog reported here is based on just the first year of the \srg\ all-sky survey, which is characterized by a duty cycle of more than 97\%.

\begin{table*}
\caption{Cross-match of the point sources from the \srg/\art\ first-year 4--12\,keV all-sky survey with selected X-ray and gamma-ray catalogs.}
\label{tab:cross_match}
\centering

\begin{tabular}{ccrrr}
\hline
X-ray survey & Energy band & Reference & Cross-  & Spurious \\
             &             &           & matches & matches \\
\hline
\rosat\ (2RXS) 1 year & 0.1--2.4\,keV & \citet{Boller16} & 478 & $\sim 2$ \\
\xmm\ SL2 & 0.2--12\,keV & \citet{Saxton08} & 454 & $\lesssim 1$ \\
MAXI/GSC 7 years  & 4--10\,keV & \citet{Kawamuro18,Hori18} & 148  & $\lesssim 3$ \\
\swift/BAT 105 months & 14--195\,keV & \citet{Oh18} & 
458 & $\sim 1$ \\
\integral\ 17 years & 17--60\,keV & Krivonos et al., (in prep.) & 345 & $\lesssim 1$ \\
\fermi/LAT 10 years & 50 MeV--1 TeV & \citet{2020ApJS..247...33A} & 88  & $\lesssim 4$ \\
\hline
\end{tabular}
\end{table*}

\subsection{Ecliptic poles}
\label{s:ecl}

The regions around the ecliptic poles receive a far higher exposure than the rest of the sky. They may therefore be regarded as deep extragalactic surveys within the \art\ all-sky survey. It is worthwhile to discuss the statistics of the detected X-ray sources separately for these zones. Specifically, we focus on the regions $b_{\rm ecl}>82^\circ$ and $b_{\rm ecl}<-82^\circ$ (with a total area of $\approx 200$\,square degrees) around the north and south ecliptic poles (SEP), respectively.

The NEP region contains 18 \art\ sources, with fluxes ranging between $\sim 2\times 10^{-13}$\,\flux\ (blazar SRGA\,J180147.2+663835=RX\,J1801.7+6638, at just 0.2\arcsec\ from the NEP) and $\sim 2\times 10^{-11}$\,\flux. All of these X-ray sources have been known before, and the majority of them are AGN, except for two clusters of galaxies and one Galactic star. The SEP region contains 37 \art\ sources, with fluxes ranging between $\sim 5\times 10^{-13}$\,\flux\ and $\sim 2\times 10^{-10}$\,\flux. A significant fraction of these sources, namely 13 (9 HMXBs and 4 SNRs), belong to the LMC. In addition to these, there are 18 AGN (confirmed or suspected), one cluster of galaxies, one Galactic HMXB, one star, and three unidentified sources (previously known 1RXS\,J061617.6$-$705237 and two newly discovered X-ray sources). 

Because of the detection threshold we chose ($S/N=4.82$, as for the rest of the sky), we expect $\approx 0.5$ spurious sources to be present in each of the NEP and SEP samples. This means that these samples should be highly pure, and it is possible to further lower the detection threshold near the ecliptic poles. We plan to do this in future work. 

\section{Summary}

We have presented the catalog of sources detected by the \art\ telescope during the first year of the ongoing \srg\ all-sky survey. It comprises 867 sources detected on the combined map of the first two half-year surveys in the 4--12\,keV energy band. The achieved sensitivity to point sources after the first year of the survey is between $\sim 4\times 10^{-12}$\,\flux\ near the ecliptic plane and $\sim 8\times 10^{-13}$\,\flux\ (4--12\,keV) near the ecliptic poles. The typical depth of the \art\ survey is comparable to that achieved in a similar energy band (4--10\,keV) in the recent \maxi/GSC all-sky survey \citep{Hori18,Kawamuro18} and is slightly worse than the sensitivity of the \xmm\ Slew Survey in the 2--12\,keV band \citep{Saxton08}. However, the \art\ survey greatly improves on the former in terms of angular resolution and provides full and well-behaved sky coverage in contrast to the latter. 

Extragalactic sources (mostly AGN and clusters of galaxies) somewhat dominate Galactic sources among the sources of known nature in the catalog. For 114 sources, \art\ has detected X-rays for the first time. Although the majority of these ($\sim 80$) are expected to be spurious, there can be a significant number of newly discovered astrophysical objects. We have started a program of optical follow-up observations of the new and previously unidentified sources, which has already led to the identification of several AGN and CVs \citep{2021AstL...47...71Z,2021arXiv210705611Z}. The expected fraction of spurious sources in the catalog drops from 10\% at the significance level $S/N\approx 4.8$ to just 1\% at $S/N\approx 5.3$, which can be taken into account in planning follow-up activities on new \art\ sources.

With the \srg/\art\ all-sky survey planned to continue for a total of four\,years (until December 2023), we can expect the number of detected sources to be growing with the completion of each next half-year scan of the sky. The number of newly discovered sources is expected to increase particularly quickly. We plan to regularly update the \art\ source catalog. The next release will be based on the data of the first two years of the survey (December 2019 -- December 2021) and will include additional information on source fluxes in various energy bands and different \srg/\art\ scans of the sky.

\section*{Acknowledgements}
The Mikhail Pavlinsky \art\ telescope is the hard X-ray instrument on board the \srg\ observatory, a flagship astrophysical project of the Russian Federal Space Program realized by the Russian Space Agency in the interests of the Russian Academy of Sciences. The \art\ team thanks the Russian Space Agency, Russian Academy of Sciences, and State Corporation Rosatom for the support of the \srg\ project and \art\ telescope. We thank Lavochkin Association (NPOL) with partners for the creation and operation of the \srg\ spacecraft (Navigator). We thank Acrorad Co., Ltd. (Japan), which manufactured the CdTe dies, and Integrated Detector Electronics AS -- IDEAS (Norway), which manufactured the ASICs for the X-ray detectors. This research was supported by grant 19-12-00396 from the Russian Science Foundation. We thank the referee for the useful comments.


\bibliographystyle{aa} 
\bibliography{art_allsky}

\begin{appendix} 

\section{Notes on individual sources}
\label{s:notes}

Here, we provide comments on the identification and classification of some sources in the \art\ catalog, namely, those with dubious identification or classification and recently discovered sources. In addition to the references provided for individual sources, we note that the matches with \xmm\ sources for a number of objects were found using the \xmm\ slew survey (XMMSL2, \citealt{Saxton08}) and the fourth \xmm\ serendipitous source catalogue \citep{Webb2020}.

\subsection*{ SRGA J000132.7+240242 }
Associated with 2MASX J00013232+2402304. Blazar candidate based on radio and infrared properties \citep{2019ApJS..242....4D}.
\subsection*{ SRGA J001315.3+774824 }
A blazar candidate \citep{2019ApJS..242....4D,2020ApJ...901....3I}, but also classified as Seyfert 2 \citep{2010A&A...518A..10V}.
\subsection*{ SRGA J002202.7+254008 }
A Seyfert 1.2 at $z=0.1292$ \citep{2017MNRAS.468..378S}.
\subsection*{ SRGA J004144.3+413415 }
Likely an LMXB in the globular cluster Bol 45 in M31 \citep{2011A&A...534A..55S}.
\subsection*{ SRGA J004241.1+411603 }
The central region of M31, unresolved into individual X-ray sources.
\subsection*{ SRGA J004506.7+620747 }
Associated with the galaxy LEDA 2631101, likely an AGN based on infrared (WISE) colors.
\subsection*{ SRGA J004546.3+413951 }
An X-ray binary in the globular cluster Bo 375 in M31, likely with a neutron star \citep{2008ApJ...689.1215B,2016MNRAS.458.3633M}.
\subsection*{ SRGA J005456.3--722646 }
In the SMC.
\subsection*{ SRGA J005642.5+604302 }
A Be star.
\subsection*{ SRGA J011511.9+882912 }
Likely associated with the bright star Gaia EDR3 576260267526450048, a spectroscopic binary in a young stellar association \citep{2020A&A...637A..43K}. Hence, likely a coronally active star.
\subsection*{ SRGA J011704.7--732637 }
In the SMC.
\subsection*{ SRGA J015639.0--835828 }
Likely associated with the star Gaia EDR3 4617143036371460864 at $D\sim 500$ pc. Possibly a CV based on X-ray and optical photometric properties.
\subsection*{ SRGA J022235.8+250816 }
A changing-look AGN \citep{2019MNRAS.484.4507V}.
\subsection*{ SRGA J022624.9+592741 }
A radio galaxy with unknown AGN optical type and redshift  \citep{2021MNRAS.500.3111B}.
\subsection*{ SRGA J022745.0--693133 }
A giant star at $D\sim 240$ pc. The exact nature of the X-ray emission is unknown.
\subsection*{ SRGA J025234.3+431004 }
Associated with the galaxy LEDA 90641, with infrared colors (WISE) typical of AGN.
\subsection*{ SRGA J030838.1--552041 }
Associated with the galaxy LEDA 410289. Present in the catalogue of quasars and active nuclei, 13th edition \citep{2010A&A...518A..10V}, without detailed classification.
\subsection*{ SRGA J031102.9--440232 }
Likely a blazar \citep{2019A&A...632A..77C}.
\subsection*{ SRGA J031131.8--315251 }
A CV (nova-like and/or polar, \citealt{2003A&A...404..301R}).
\subsection*{ SRGA J035023.8--501802 }
A Seyfert 2 \citep{2017ApJ...850...74K}.
\subsection*{ SRGA J040850.8--791418 }
Likely associated with the star Gaia EDR3 4625832751643853696 at $D\sim 1300$ pc. Possibly a CV based on X-ray and optical photometric properties.
\subsection*{ SRGA J041328.3--061446 }
A flat-spectrum radio source \citep{2007ApJS..171...61H}.
\subsection*{ SRGA J041732.7--525316 }
Likely associated with the galaxy LEDA 14813.
\subsection*{ SRGA J042616.9--592322 }
Associated with the galaxy LEDA 371722, with infrared (WISE) colors typical of AGN.
\subsection*{ SRGA J043209.5+354927 }
A Seyfert 1 at $z=0.0506$ \citep{2021AstL...47...71Z}.
\subsection*{ SRGA J043510.5--752749 }
Likely associated with CRTS J043509.7--752743, a variable star of RR Lyr type. The exact nature of the X-ray emission is unknown.
\subsection*{ SRGA J043522.9+552234 }
X-ray transient SRGA J043520.9+552226 = SRGE J043523.3+552234 discovered by {\it SRG}/ART-XC and eROSITA, associated with the optical transient ATLAS19bcxp. An LMXB, BH candidate \citep{2020ATel13571....1M,2020arXiv201200169Y,2021arXiv210705588M}.
\subsection*{ SRGA J043830.9--681205 }
Likely associated with the galaxy 2MASS J04383119--6812003, with infrared colors (WISE) typical of AGN.
\subsection*{ SRGA J045050.1+301450 }
A Seyfert 1.9 at $z=0.0331$ \citep{2021AstL...47...71Z}.
\subsection*{ SRGA J045254.2--585353 }
A blazar candidate based on radio and infrared properties \citep{2019ApJS..242....4D}.
\subsection*{ SRGA J050021.6+523801 }
Associated with the gamma-ray source 3FGL J0500.3+5237, likely a blazar \citep{2019A&A...632A..77C}.
\subsection*{ SRGA J050810.6--660657 }
A new Be X-ray binary eRASSU J050810.4--660653 discovered by {\it SRG}/eROSITA in LMC \citep{2020ATel13609....1H}, previously detected during the ROSAT all-sky survey (2RXS J050810.2--660645).
\subsection*{ SRGA J051514.3--142701 }
Associated with the infrared source WISEA J051514.91--142718.6, with colors typical of AGN.
\subsection*{ SRGA J052029.0--715733 }
In the LMC.
\subsection*{ SRGA J052410.7--662045 }
In the LMC.
\subsection*{ SRGA J052505.4--693853 }
In the LMC.
\subsection*{ SRGA J052947.3--655645 }
In the LMC.
\subsection*{ SRGA J053043.5--665424 }
In the LMC.
\subsection*{ SRGA J053232.1--655141 }
In the LMC.
\subsection*{ SRGA J053249.3--662220 }
In the LMC.
\subsection*{ SRGA J053525.0--691621 }
In the LMC.
\subsection*{ SRGA J053739.1+210826 }
A binary Be star.
\subsection*{ SRGA J053749.8--691016 }
In the LMC.
\subsection*{ SRGA J053857.2--640503 }
In the LMC.
\subsection*{ SRGA J053939.4--694440 }
In the LMC.
\subsection*{ SRGA J054010.9--692001 }
In the LMC.
\subsection*{ SRGA J055053.7--621457 }
Associated with the galaxy LEDA 178653, with infrared colors (WISE) typical of AGN.
\subsection*{ SRGA J060241.1--595152 }
Associated with the galaxy LEDA 178859, with infrared colors (WISE) typical of AGN.
\subsection*{ SRGA J060651.8--624550 }
Associated with the galaxy LEDA 340165, with infrared colors (WISE) typical of AGN.
\subsection*{ SRGA J060728.9--614836 }
Heavily obscured AGN \citep{2009MNRAS.398.1165G,2020MNRAS.497..229A} in an edge-on galaxy.
\subsection*{ SRGA J061322.9--290027 }
Associated with the galaxy LEDA 734640, with infrared (WISE) colors typical of AGN.
\subsection*{ SRGA J061619.2--705228 }
Possibly associated with the infrared source WISEA J061617.13--705228.7 based on the position of the X-ray source 4XMM J061617.1-705229.
\subsection*{ SRGA J062109.8--680554 }
Associated with the galaxy LEDA 179145, with infrared (WISE) colors typical of AGN.
\subsection*{ SRGA J062627.2+072734 }
Associated with the galaxy LEDA 136513, with infrared (WISE) colors typical of AGN.
\subsection*{ SRGA J062945.0--834426 }
Likely associated with the infrared source WISEA J062948.62--834421.6, with colors typical of AGN.
\subsection*{ SRGA J062953.4--033505 }
Likely associated with the star Gaia EDR3 3105034928633782400 at $D\sim 670$\,pc. Possibly a CV based on X-ray and optical photometric properties.
\subsection*{ SRGA J063558.6+075528 }
Associated with the bright star TYC 733-2098-1 at $D\sim 200$ pc. Coronally active or a CV?
\subsection*{ SRGA J064849.9--694524 }
A blazar candidate \citep{2019A&A...632A..77C}.
\subsection*{ SRGA J065513.5--012846 }
Likely associated with the bright star V520 Mon (long-period variable) at $D\sim 4000$ pc. Possibly an HMXB or a symbiotic X-ray binary.
\subsection*{ SRGA J070636.4+635109 }
Associated with the galaxy UGC 3660. Classified as a Seyfert 1 based on follow-up spectroscopy (Zaznobin et al., in preparation).
\subsection*{ SRGA J071739.0--710349 }
Associated with the infrared source WISEA J071740.40--710347.3, with colors typical of AGN.
\subsection*{ SRGA J072041.5--552614 }
Associated with the galaxy LEDA 409410, with infrared (WISE) colors typical of AGN.
\subsection*{ SRGA J072823.5--440823 }
Associated with the galaxy 2MASS J07282338--4408241, with infrared colors (WISE) typical of AGN.
\subsection*{ SRGA J080735.3+023517 }
Likely associated with the star Gaia EDR3 3090944824560398208 at  $D\sim 1400$ pc. Possibly a CV based on X-ray and optical photometric properties.
\subsection*{ SRGA J082625.5--703138 }
Likely a nonmagnetic CV \citep{2012A&A...545A.101P}.
\subsection*{ SRGA J084434.0--375747 }
Likely associated with the bright star HD 74771 at $D\sim 170$ pc. Coronally active?
\subsection*{ SRGA J085040.6--421156 }
Likely associated with the bright star UCAC2 13726137 at $D\sim 12$ kpc. Possibly an HMXB or a symbiotic X-ray binary.
\subsection*{ SRGA J091511.9--752345 }
Associated with the galaxy 2MASS J09151520--7523498, with infrared colors (WISE) typical of AGN. Also a radio source (SUMSS).
\subsection*{ SRGA J092021.6+860249 }
Associated with the galaxy LEDA 2790304, with infrared (WISE) colors typical of AGN; a radio source (NVSS).
\subsection*{ SRGA J092418.4--314218 }
An LMXB or a CV \citep{2017ApJS..230...25T}.
\subsection*{ SRGA J104450.9--602446 }
Likely an AGN \citep{2019ApJ...887...32C}.
\subsection*{ SRGA J104833.4--390227 }
A Seyfert 1.5 at $z=0.0446$ \citep{2018ATel11414....1M}.
\subsection*{ SRGA J111153.2--611826 }
A massive star-forming region. Very many point X-ray sources and diffuse X-ray emission ({\it Chandra}, \citealt{2014ApJS..213....1T}).
\subsection*{ SRGA J111457.9--611432 }
A massive star-forming region. Very many point X-ray sources and diffuse X-ray emission ({\it Chandra}, \citealt{2014ApJS..213....1T}).
\subsection*{ SRGA J111821.7--543730 }
Likely an LMXB and not an HMXB \citep{2013A&A...560A.108C}.
\subsection*{ SRGA J114721.9--495309 }
Possible confusion of two known X-ray sources: XMMSL2 J114720.7--495302 (14\arcsec\ from the ART-XC position) and XMMSL2 J114724.7--495303 (27\arcsec, the brighter one). There is also 1RXS J114724.3--495250 at 29\arcsec. The weaker XMM source is associated with the emission-line star TWA 19B, while the brighter XMM source is associated with the young stellar object HD 102458.
\subsection*{ SRGA J120413.4--294646 }
Likely associated with the infrared source WISEA J120412.71--294709.1, with colors typical of AGN.
\subsection*{ SRGA J123629.1-664554 }
In addition to XMMSL2 J123629.8--664549, there is another nearby source 2SXPS J123632.4--664557.
\subsection*{ SRGA J123821.8--253205 }
A bright, short-duration X-ray transient SRGt J123822.3--253206 discovered by ART-XC and eROSITA \citep{2020ATel13415....1S,2020ATel13416....1W}.
\subsection*{ SRGA J124249.6-630348 }
A Be star or Be star--white dwarf binary \citep{2018PASJ...70..109T}.
\subsection*{ SRGA J124640.6+543222 }
A Seyfert 2 \citep{2012A&A...545A.101P}.
\subsection*{ SRGA J131239.5--624256 }
A Wolf-Rayet star. The most likely physical picture is that of colliding stellar winds in a wide binary system, with the unseen secondary star being another WR star or a luminous blue variable 
\citep{2014ApJ...785....8Z}.
\subsection*{ SRGA J132032.4--701437 }
Associated with the bright star Gaia EDR3 5844075864125790464 at $D\sim 2300$ pc. Possibly an HMXB or a symbiotic X-ray binary.
\subsection*{ SRGA J133949.9--643019 }
Possibly associated with IGR J13402--6428 at 2.8\,arcmin from the ART-XC position, but the latter is inconsistent with those of two {\it Chandra} sources, CXOU J133935.8--642537 and CXOU J133959.2--642444 (located at 5.0 and 4.3 arcmin, respectively, from the {\it INTEGRAL} position), which have been suggested as possible soft X-ray counterparts of IGR J13402--6428 \citep{2012ApJ...754..145T}.
\subsection*{ SRGA J140654.9--52749 }
Associated with the galaxy NGC 5472, a radio source.
\subsection*{ SRGA J141249.5--402141 }
A CV of VY Scl-type \citep{2010AN....331..227G}.
\subsection*{ SRGA J152100.6+320405 }
A Seyfert 2 at $z=0.1143$ \citep{2021AstL...47...71Z}.
\subsection*{ SRGA J153414.5+625852 }
A Seyfert 1 at $z=0.26$ \citep{2015ATel.7177....1C}.
\subsection*{ SRGA J153814.0--554219 }
Likely an LMXB \citep{2012A&A...540A..22D}.
\subsection*{ SRGA J154805.3--473806 }
Associated with the galaxy LEDA 141863, with infrared colors (WISE) typical of AGN.
\subsection*{ SRGA J160050.0--514249 }
A colliding-wind Wolf-Rayet binary \citep{2019NatAs...3...82C}.
\subsection*{ SRGA J160456.1--722318 }
Likely associated with the star Gaia EDR3 5806631686386070144 at $D\sim 580$ pc. Possibly a CV based on X-ray and optical photometric properties.
\subsection*{ SRGA J160901.4--390519 }
Likely associated with the star THA 15-35 at $D\sim 160$ pc.
\subsection*{ SRGA J161017.3--634243 }
A Seyfert 1.5, at $z=0.2094$ \citep{2018ATel11341....1S}.
\subsection*{ SRGA J161156.7--603826 }
Part of cluster A3627, together with SRGA J161415.5--605124.
\subsection*{ SRGA J161201.8--464616 }
Likely associated with the star Gaia EDR3 5990098842316868352 at $D\sim 400$ pc. Possibly a CV based on X-ray and optical photometric properties.
\subsection*{ SRGA J161415.5--605124 }
Part of cluster A3627, together with SRGA J161156.7--603826.
\subsection*{ SRGA J163529.2--480558 }
Bright star at $D\sim 1100$ pc. Possibly an HMXB or a symbiotic X-ray binary.
\subsection*{ SRGA J174009.6--284714 }
Likely a CV rather than an LMXB \citep{2014ApJ...786...20L}.
\subsection*{ SRGA J174028.2--365545 }
An intermediate polar \citep{2019ApJ...887...32C}.
\subsection*{ SRGA J174046.4+060353 }
A CV \citep{2015AJ....150..170H}.
\subsection*{ SRGA J174446.1--295057 }
Symbiotic X-ray binary \citep{2014MNRAS.441..640B}.
\subsection*{ SRGA J174450.9--323321 }
Massive cluster of galaxies at $z=0.055$, with some contribution from a background blazar in the hard X-ray band \citep{2015ApJ...799...24B}.
\subsection*{ SRGA J175721.3--304400 }
Likely an LMXB with a giant companion \citep{2013AdSpR..51.1278M}.
\subsection*{ SRGA J180849.4+663432 }
A flat-spectrum radio source \citep{2014ApJS..213....3M}.
\subsection*{ SRGA J181228.0--181235 }
A burster, an ultra-compact LMXB \citep{2019MNRAS.486.4149G}.
\subsection*{ SRGA J181239.6--221924 }
Discovered by MAXI, likely an LMXB \citep{2018ATel12254....1N}.
\subsection*{ SRGA J181637.1--391248 }
Likely associated with V371 CrA, a variable star of Mira Cet type. A symbiotic X-ray binary?
\subsection*{ SRGA J182156.2+642036 }
Powerful AGN in the central galaxy of a rich cluster. The X-ray luminosities of the quasar and cluster are comparable according to previous observations, in particular, {\it Chandra} \citep{2010MNRAS.402.1561R,2014ApJ...792L..41R}.
\subsection*{ SRGA J182511.1+645018 }
KO giant star with blended double lines in the optical spectrum \citep{2019A&A...626A..31S}, thus most likely a coronally active binary.
\subsection*{ SRGA J182919.6--121301 }
Likely an intermediate polar \citep{2016MNRAS.461..304C}.
\subsection*{   SRGA J183605.4--192217 }
Possibly associated with 3MAXI J1836--194 at 2.0 arcmin from the ART-XC position.
\subsection*{ SRGA J183754.6+155438 }
Likely associated with the bright star Gaia EDR3 4510001297611019392 at $D\sim 500$ pc.
\subsection*{ SRGA J183907.1--571456 }
Suggested to be associated with the infrared source WISEA J183905.95--571505.1, with colors typical of AGN, and also has an absorbed X-ray spectrum \citep{2017MNRAS.470.1107L}. However, the optical counterpart appears to have a high proper motion in Gaia (Gaia EDR3 6637527637032850176).
\subsection*{ SRGA J184117.8--045612 }
Magnetar 1E 1841--045 in the supernova remnant Kes 73, which are both bright in X-rays \citep{2013ApJ...779..163A,2014ApJ...781...41K} and consistent with the ART-XC position.
\subsection*{ SRGA J184554.7--003937 }
A Be X-ray binary, pulsar \citep{2019ATel13195....1K,2019ATel13211....1M}.
\subsection*{ SRGA J185421.6+083844 }
Likely associated with the infrared source CatWISE J185422.29+083846.1, with infrared colors typical of AGN.
\subsection*{ SRGA J190140.2+012627 }
Likely an LMXB \citep{2008AstL...34..753K,2010A&A...510A..61T}.
\subsection*{ SRGA J190722.0--204635 }
Likely an intermediate polar \citep{2019MNRAS.489.3031X}.
\subsection*{ SRGA J191456.9+103647 }
Likely an HMXB \citep{2015MNRAS.446.1041C}.
\subsection*{ SRGA J192501.3+504309 }
A Seyfert 1.2 at $z=0.068$ \citep{2013A&A...556A.120M}.
\subsection*{ SRGA J194638.9+704552 }
A CV, likely an intermediate polar \citep{2021arXiv210705611Z}.
\subsection*{ SRGA J195702.4+615036 }
Associated with the galaxy LEDA 2625686, with infrared (WISE) colors typical of AGN.
\subsection*{ SRGA J195928.3+404358 }
There must be a significant contribution of the central AGN to the X-ray luminosity of the cluster \citep{2002ApJ...564..176Y}.
\subsection*{ SRGA J201118.2+600421 }
Suggested to be an AGN \citep{2012A&A...548A..99W}, but likely associated with the star Gaia EDR3 2236896418906579072 at $D\sim 1.5$ kpc  based on the position of XMMSL2 J201116.8+600431. Possibly a CV based on X-ray and optical photometric properties.
\subsection*{ SRGA J202932.4--614903 }
Associated with the galaxy LEDA 352109, with infrared (WISE) colors typical of AGN.
\subsection*{ SRGA J204149.6--373345 }
Blazar candidate \citep{2019A&A...632A..77C} or a cluster of galaxies \citep{2000AN....321....1S}.
\subsection*{ SRGA J204319.7+443821 }
Galactic X-ray transient SRGA J204318.2+443815 = SRGE J204319.0+443820 discovered by {\it SRG}/ART-XC and eROSITA \citep{2020ATel14206....1M}, an X-ray pulsar in a Be system \citep{2021arXiv210705587L}.
\subsection*{ SRGA J204547.8+672642 }
A CV \citep{2021arXiv210705611Z}.
\subsection*{ SRGA J211747.6+513850 }
Blazar candidate based on infrared (WISE) colors \citep{2013ApJS..206...17M}.
\subsection*{ SRGA J221913.2+362014 }
Associated with the infrared source WISEA J221914.50+362010.5, with colors typical of AGN; a radio source (NVSS).
\subsection*{ SRGA J223714.9+402939 }
Associated with the galaxy LEDA 5060459, with infrared (WISE) colors typical of AGN.
\subsection*{ SRGA J225412.8+690658 }
A CV \citep{2021arXiv210705611Z}.
\subsection*{ SRGA J230630.9+155637 }
Apparently a Seyfert 2 based on a visual inspection of the SDSS DR16 spectrum.
\subsection*{ SRGA J232037.8+482329 }
Associated with the galaxy LEDA 2316409, a radio source.
\subsection*{ SRGA J235250.6--170449 }
Associated with the galaxy 2MASS J23525142--1704372, with infrared colors (WISE) typical of AGN.

\end{appendix}

\section{Source catalog}
\label{s:sources}

In Table~\ref{tab:longtable}, we present the catalog of sources detected during the first year of the \srg/\art\ all-sky survey.

\longtab[1]{
\begin{landscape}
\label{tab:longtable}
\begin{longtable}{rrrrcccrrrrrr}
\caption{Catalog of sources\tablefootmark{a)} detected during the first year of the \srg/\art\ all-sky survey.}\\
\hline
Id & Name & RA & Dec & S/N & Flux & Common name & Redshift & Class  \\
&  SRGA & (J2000)& (J2000) & & ($10^{-12}$\,\flux)  & & &   \\
\hline
\endfirsthead
\caption{Continued.} \\
\hline
Id & Name & RA & Dec & S/N & Flux & Common name & Redshift & Class  \\
& SRGA & (J2000)& (J2000) && ($10^{-12}$\,\flux)  & & &   \\
\hline
\endhead
\hline
\endfoot
\hline
\endlastfoot
N001&{\bf  J000132.7+240242 }&0.3861&24.0449&5.3&$4.7 (-2.3 +3.1)$&&0.1045&BLAZAR?$^{c)}$\\
N002& J000632.2--690029&1.6340&-69.0081&7.1&$5.0 (-2.0 +2.6)$&CTCV J0006-6900&&CV\\
N003& J001124.6--112841&2.8525&-11.4782&5.3&$8.3 (-2.5 +3.0)$&WW Cet&&CV\\
N004& J001315.3+774824&3.3138&77.8066&5.3&$4.6 (-1.5 +1.8)$&8C 0010+775&0.3260&BLAZAR?$^{c)}$\\
N005& J001704.0+813451&4.2667&81.5810&7.3&$3.8 (-1.4 +1.7)$&6C 001403+811827&3.3660&BLAZAR\\
N006&{\bf  J002021.4--042502 }&5.0893&-4.4171&5.2&$2.8 (-2.1 +3.0)$&&&UNIDENT\\
N007& J002202.7+254008&5.5113&25.6689&5.1&$4.4 (-2.2 +2.9)$&XMMSL1 J002202.9+254004&0.1292&SEYFERT$^{c)}$\\
N008& J002256.7+614104&5.7362&61.6845&6.2&$4.4 (-1.6 +1.9)$&V1033 Cas&&CV\\
\end{longtable}
\tablefoot{ 
\tablefoottext{a}{For the description of columns see \S\ref{s:catalog}. Online version of this catalog is available at \url{http://srg.cosmos.ru}.}
\tablefoottext{b}{Spatial extension of the source is detected.}
\tablefoottext{c}{Additional information on the source identification/classification is provided in Sect.~\ref{s:sources}.}
}
\end{landscape}
}

\end{document}